\documentclass[a4paper,11pt,headings=big,DIV=12]{scrartcl}
\pdfoutput=1
\usepackage{amsmath,amssymb,mathtools}
\usepackage[table]{xcolor}
\usepackage{tcolorbox}
\definecolor{mygray}{gray}{0.2}
 \addtokomafont{disposition}{\rmfamily\boldmath}
  \usepackage{bbold}
   \usepackage{environ}

\NewEnviron{thincases}{\scalebox{0.5}[1]{$\displaystyle
\left\{\scalebox{2}[1]{\setlength{\arraycolsep}{6pt}
$\displaystyle\begin{array}{ll}
\BODY
\end{array}$}\right.$}}

\definecolor{mypink1}{rgb}{0.9, 0.2, 0.6}
\usepackage{graphicx}
\addtokomafont{disposition}{\rmfamily\boldmath}
\usepackage[utf8]{inputenc}
\usepackage{enumerate}
\usepackage{slashed}
\usepackage{cite}
\usepackage{comment}
\usepackage{multirow}
\usepackage{acronym}
 \usepackage{bbold}
 \usepackage{tikz-feynman}
 \usepackage{jheppub}
\allowdisplaybreaks
\numberwithin{equation}{section}

\newcommand{\Ima}{\textrm{Im}}

\newcommand{\vev}[1]{\langle #1 \rangle}

\newcommand{\matel}[3]{\langle #1|#2|#3\rangle}

\newcommand{\be}{\beta}
\newcommand{\ga}{\gamma}
\newcommand{\Ga}{\Gamma}
\newcommand{\de}{\delta}
\newcommand{\De}{\Delta}
\newcommand{\la}{\lambda}

\newcommand{\eps}{\epsilon}

\newcommand{\sig}{ \sigma}
\newcommand{\Sig}{ \Sigma}

\newcommand{\GeV}{\,\mbox{GeV}}
\newcommand{\MeV}{\,\mbox{MeV}}

\newcommand{\MSbar}{\overline{\text{MS}}}

\newcommand{\EQ}{Eq.~}
\newcommand{\EQs}{Eqs.~}
\newcommand{\TAB}{table~}

\newcommand{\FIG}{Fig.~}

\newcommand{\SEC}{section~}

\newcommand{\APP}{appendix~}

\newcommand{\REF}{ref.~}
\newcommand{\REFs}{refs.~}

\newcommand{\Fone}{{v}}
\newcommand{\Lag}{{\cal L}}

\newcommand{\gone}{A}
\newcommand{\gtwo}{ D}

\newcommand{\gJ}{J}
\newcommand{\ferm}{N}

\newcommand{\Lageff}{\Lag_{\text{eff}} }

\newcommand{\XX}{\chi}

\newcommand{\gast}{\ga_*}

\newcommand{\LsM}{linear $\sig$-model }

\newcommand{\party}{\varphi}
\newcommand{\mom}{\Omega}
\newcommand{\form}{G}
\newcommand{\back}{b}

\newcommand{\ORD}{{\cal O}}

\newcommand{\Tud}[2]{T^{#1}_{\;\; #2}}
\newcommand{\TEMT}{ \Tud{\rho}{\rho} }

\newcommand{\mink}{\eta}

\newcommand{\Op}{{\cal O}}

\newcommand{\GaC}{\Theta}

\newcommand{\rE}[1]{r_{\mathrm{E},#1}}
\newcommand{\mE}[1]{m_{\mathrm{E},#1}}

\newcommand{\Mass}{\zeta}
\newcommand{\meff}{  m_{ \mathrm{eff}} }
\newcommand{\reff}{  r_{\mathrm{eff}}  }
\newcommand{\res}[2]{r_{#1}^{#2}}

\definecolor{violet}{rgb}{0.94, 0.2, 0.8}
\definecolor{lightblue}{rgb}{0.39, 0.58, 0.93} 
\definecolor{lightgreen}{rgb}{0.1, 0.73, 0.33}
\newcommand{\comb}[1]{
{#1}}

\newcommand{\com}[1]{
{#1}}

\DeclareOldFontCommand{\tt}{\normalfont\ttfamily}{\mathtt}
\DeclareOldFontCommand{\bf}{\normalfont\bfseries}{\mathbf}

\begin{document}

\title{\boldmath  Gravitational $D$-Form Factor:  The $\sigma$-Meson \\
as a Dilaton confronted  with  Lattice QCD Data I  }

\author[1]{Roy Stegeman,}
\author[1]{Roman Zwicky,}

\affiliation[1]{Higgs Centre for Theoretical Physics, School of Physics and
Astronomy, The University of Edinburgh, 
Peter Guthrie Tait Road, Edinburgh EH9 3FD, Scotland, UK}
\emailAdd{r.stegeman@ed.ac.uk}
\emailAdd{roman.zwicky@ed.ac.uk}

\abstract{
We investigate the nucleon and pion gravitational $D$-form factors, by fitting a
$\sigma/f_0(500)$-meson pole, together with a background term,  to lattice data
at  $m_\pi \approx 170\MeV$. We  find that the fitted residues are compatible with predictions from dilaton
effective theory. In this framework, the $\sigma$-meson takes on the role of the dilaton,
the Goldstone boson of spontaneously broken scale symmetry.
These results support  the idea that QCD may be governed by an infrared fixed point and offer a physical interpretation of the $D$-form factor (or $D$-term)  in the soft limit.}

\maketitle

\section{Introduction}

Gravitational form factors probe the energy-momentum tensor for physical states, such as nucleons, through matrix elements of the form $\matel{N(p')}{T_{\mu\nu}}{N(p)}$.
The momentum transfer $q = p' -p$ reveals  the energy-momentum distribution of the nucleon just like electromagnetic
form factors test the charge distribution.
Defined a long time ago \cite{Kobzarev:1962wt,Pagels:1966zza}, gravitational form factors have seen a growing interest (reviewed in \cite{Polyakov:2018zvc,Burkert:2023wzr,Lorce:2024ipy,Lorce:2025oot}).
In part related to their    experimental accessibility, through the first
moment of  generalised parton distributions \cite{Ji:1996ek},  in  deeply virtual Compton scattering
 \cite{Burkert:2018bqq,Duran:2022xag,Goharipour:2025lep}, near-threshold $J/\psi$ photoproduction \cite{GlueX:2023pev}, and  
 $\ga^*\ga \to \pi\pi$
  \cite{Diehl:1998dk,Belle:2015oin,Kumano:2019vlv},   among others.  This has triggered  lattice QCD investigations
\cite{Shanahan:2018pib,Pefkou:2021fni,Hackett:2023rif,Hackett:2023nkr,Wang:2024lrm,Abbott:2025irb}
 which provide  the basis of this paper.
  In addition to experimental 
and lattice results,  there are perturbative approaches  in chiral theories at low $q^2$ \cite{Belitsky:2002jp,Alharazin:2023zzc} and  light-cone sum rules
at high $q^2$ \cite{Anikin:2019kwi,Tong:2021ctu,Tong:2022zax,Ozdem:2022zig,Dehghan:2025ncw},
dispersive analyses \cite{Broniowski:2024oyk,Cao:2024zlf,Broniowski:2025ctl,Cao:2025dkv},
   Skyrme-based models \cite{Cebulla:2007ei,Jung:2013bya,Tanaka:2025pny,Fukushima:2025jah}, light-front quark models \cite{Chakrabarti:2020kdc,Choi:2025rto},
   and holographic models \cite{Mamo:2021krl,Mamo:2022eui,Fujita:2022jus,Sugimoto:2025btn,Tanaka:2025znc}.

Conserved-current form factors often obey physical constraints at zero momentum transfer.
An example is  the charged-pion electromagnetic form factor  $F_+(0) =1$, expressing  charge conservation.
The infrared interpretation of the  gravitational $D$-form factor, associated with the internal pressure distribution \cite{Polyakov:2002yz},  has remained elusive and  puzzled the community for a long time \cite{Polyakov:2018zvc}, as debated in  \cite{Ji:2025gsq,Ji:2025qax,Lorce:2025oot}. Our work offers an explanation in terms of the dilaton.

The idea that  strong interactions are governed by an infrared fixed point \cite{Isham:1970gz,Zumino:1970tu,Ellis:1971sa} has recently been reexamined in QCD, both in low-energy processes  \cite{Crewther:2013vea,Crewther:2015dpa} and more formally \cite{Zwicky:2023bzk,Zwicky:2023krx,Zwicky:2025moo}, by matching scaling dimensions of the underlying theory to those of the effective theory.
Consistency of the quark-mass anomalous dimension with lattice simulations \cite{DelDebbio:2015byq,Hasenfratz:2020ess,Fodor:2017gtj,LatticeStrongDynamics:2018hun,LatticeStrongDynamics:2023bqp,Bennett:2024tex,LatKMI:2025kti}
(or fits thereto  \cite{Appelquist:2017wcg,Appelquist:2017vyy}),
phenomenological models,
lower dimensional models \cite{Cresswell-Hogg:2025kvr}
and  ${\cal N} =1 $
supersymmetric gauge theories  \cite{Terning:2006bq,Zwicky:2023krx,Shifman:2023jqn} are established.
The underlying idea of this scenario is that spontaneous scale symmetry breaking generates  hadron masses,
with the resulting Goldstone bosons, the dilaton, realising the corresponding
Ward identities.\footnote{Gauge theory dilatons are of interest elsewhere:  the
Higgs as a dilaton \cite{Matsuzaki:2012xx,Dietrich:2005jn,Cata:2018wzl,Zwicky:2023krx} or  nuclear physics in dense 
matter  \cite{Brown:1991kk,Rho:2021zwm,Rho:2024jdo}.}

The  gravitational form factors for a scalar $\varphi$
provide the ideal setting to illustrate these ideas
\begin{equation}
\label{eq:GFF}
\GaC_{\mu\nu}(q)  \equiv  \matel{\varphi(p')}{T_{\mu\nu}(0)}{\varphi(p)} =
   2 {\cal P}_\mu {\cal P}_\nu  A (q^2) +
\frac{1}{2} (q_\mu q_\nu-  q^2 \eta _{\mu\nu} )  D (q^2)  \;,
\end{equation}
with  momenta $  2{\cal P} \equiv p + p'$ and $q$ as defined above.
 Energy conservation implies the model-independent constraint  $ A(0) =1 $
 and if one assumes that $D(q^2)$ is regular for $q \to 0$,
    one recovers the standard textbook formula
 $  \GaC(0) \equiv \GaC^{\rho}_{\phantom{\rho} \rho}(0) = 2 m_\varphi^2$  \cite{Donoghue:1992dd}
 {(for the normalisation $\vev{\varphi(p')|\varphi(p)} =  2 E_p  (2\pi)^3 \de (\vec{p}- \vec{p}\,')$)}.
However, in the presence of a massless dilaton arising from spontaneous scale symmetry breaking, the dilaton pole
  \begin{equation}
 \label{eq:Golberger}
  D(q^2) =  \frac{4}{3} \frac{m_\varphi^2}{q^2} +  \ORD(1) \;,
   \end{equation}
 modifies  the textbook formula  to satisfy the infrared conformal Ward identity ``$\TEMT =0$",
  analogous to the Goldberger-Treiman mechanism, where the pion
  restores  the chiral Ward identity \cite{Donoghue:1992dd}.\footnote{In the axial singlet channel, the $\eta'$  plays a similar role in QCD processes, see for instance \cite{Tarasov:2020cwl,Tarasov:2021yll,Tarasov:2025mvn,Bhattacharya:2022xxw,Bhattacharya:2023wvy}. 
For the dilaton it was  proposed  in \REF\cite{Gell-Mann:1969rik}, and verified using the LSZ formalism~ \cite{DelDebbio:2021xwu}
and  the effective theory~ \cite{Zwicky:2023fay}.}
 Consequently, one finds
\begin{equation}
  \GaC(0)  = \matel{\varphi(p)}{\TEMT(0)}{\varphi(p)}  =
  \begin{thincases}
2 m_\varphi^2  &   \quad \text{textbook formula} \\[-0.2cm]
 0 &   \quad \text{dilaton pole}
\end{thincases}  \;.
\end{equation}

In this work, we test infrared conformality in a direct and physical way,
independent of $\beta$-functions.
Our strategy is to fit lattice data for the nucleon and pion
gravitational $D$-form factor~\cite{Hackett:2023rif,Hackett:2023nkr}
to the predictions of leading-order (LO) dilaton effective theory~\cite{Zwicky:2023fay},
which incorporates the dilaton Goldberger-Treiman mechanism described above.

The paper is organised as follows. In \SEC\ref{sec:GFF} the  gravitational form factors are defined
for the nucleon and pion,  and the LO dilaton predictions are given.  Section \ref{sec:dom} motivates a fit ansatz suitable in the Euclidean regime.
 In \SEC\ref{sec:fits} we present our fits, test the dilaton hypothesis
 and comment on the $D$-term.
 Conclusions and discussions follow in \SEC\ref{sec:conc}. Appendix \ref{app:LsigM} explores the \LsM to clarify the relation between complex-valued and effective Euclidean poles, while \APP\ref{app:multipole} sketches a multipole expansion in momentum space to motivate the fit ansatz. {In \APP\ref{app:comparefits} we present
additional plots putting our fits into perspective with other work.}

\section{Gravitational Form Factors}
\label{sec:GFF}

\subsection{Definition of nucleon and pion form factors}

The gravitational form factors of the nucleon $N$ and the pion $\pi$
are parametrised  by\comb{
 \begin{alignat}{2}
 \label{eq:gffdefinition}
&  \Theta^N_{\mu\nu}(q)  &\;=\;&   a_{\mu\nu} \,\gone^\ferm (q^2) + j_{\mu\nu} \,\gJ^\ferm(q^2) + 
d_{\mu\nu} \,\gtwo^\ferm (q^2) \;, \nonumber  \\[0.1cm]
&  \Theta^\pi_{\mu\nu}(q) &\;=\;&  a_{\mu \nu} \,\gone^\pi (q^2) +
d_{\mu\nu} \,\gtwo^\pi (q^2) \;,
\end{alignat}
where the left hand side is defined by
\begin{equation}
 \frac{1}{2 m_\ferm} \, \bar u(p',s') \Theta^N_{\mu\nu}(q) u(p,s)  \equiv \matel{\ferm(p',s')}{T_{\mu\nu}}{\ferm(p,s)} 
\;, \quad  \Theta^\pi_{\mu\nu}(q) \equiv  \matel{\pi(p')}{T_{\mu\nu}}{\pi(p)}  \;,
\end{equation}}
and  $u(p,s)$ denotes the Dirac spinor, with ($\sig_{\mu \nu} = \tfrac{i}{2} [ \ga_\mu , \ga_\nu ]$
and  $\sig_{q\nu} = \sig_{\mu\nu} q^\mu$)
\begin{equation}
a_{\mu\nu} =2 {\cal P}_\mu {\cal P}_\nu \;, \quad d_{\mu\nu}  =\frac{1}{2} (q_\mu q_\nu-  q^2 \eta _{\mu\nu} ) \;, \quad
j_{\mu\nu} = 2 i \, {\cal P}_\mu \sig_{q \nu } + {\mu \leftrightarrow \nu } \;,
\end{equation}
Lorentz structures ensuring translational invariance  $q^\mu \GaC_{\mu\nu} (q)= 0$.
{The normalisation    $ A(0) =1 $ holds model-independently since
 $T_{\mu\nu}$ is the associated Noether current with conserved momenta $P_\mu = \int d^3 x T_{\mu 0}$.}
 Furthermore, for the nucleon one has $J^N(0) = \tfrac{1}{2}$, which reflects the absence of an anomalous gravitational magnetic moment, a consequence of the universality of
gravity.\footnote{The form factors  $A^{\pi,N}$ and $J^N$ are analogues of the electromagnetic pion, Dirac and Pauli form factors.}

\subsection{Leading order dilaton effective theory and $D(q^2)$}
\label{sec:EFT}

The form factors have been evaluated in the LO dilaton effective  theory in an arbitrary spacetime dimension $d$, including  chiral corrections,  in \REF\cite{Zwicky:2023fay}.
Here, we consider it  instructive to highlight the main effect,   the dilaton pole in the $D$-form factor,
using the nucleon as an example. Denoting the dilaton by $\sig$
and the coset field by $\hat{\chi} \equiv  e^{ - \sig/F_\sig}$,  the relevant Lagrangian reads
\begin{equation}
\label{eq:Leff}
\Lageff =  \frac{1}{2} F_\sig^2 ( (\partial \hat{ \XX})^2 -  \frac{1}{6} \,  R \, \hat{\XX}^{2}  )
\; +   \hat{\XX}^{3 - 2 \omega_N} \bar{\ferm}  (i \slashed{\Delta} - \hat{\XX} m_\ferm)  \ferm  \;,
\end{equation}
where   $F_\sig$, $R$, $\omega_N$ and   $\Delta_\mu$  stand for
  the dilaton decay constant,  the Ricci scalar,
   the nucleon Weyl weight (conformal charge) and the
 Weyl-covariant derivative, respectively.  For the latter  we refer the reader to original
\REFs  \cite{Mack:1969rr,Isham:1970gz,Isham:1970xz}, and
\cite{Zwicky:2023fay} for   an emphasis on generic Weyl-weight.
We first focus on the improvement term proportional to the Ricci scalar. While irrelevant for scattering in flat space, it contributes to the energy-momentum tensor
\begin{equation}
\label{eq:TRdef}
T_{\mu\nu} \supset   T^R_{\mu\nu} =   \frac{F_\sig}{6} (\mink_{\mu\nu} \partial^2 -  \partial_{\mu} \partial_{\nu} )   \hat{ \XX}^{2} \;,
\end{equation}
since it is an effective coupling to gravity.  It is crucial as it realises the Goldstone matrix element
in the effective theory
\begin{equation}
 \matel{0}{  T^R_{\mu\nu}  }{\sig}
=  \frac{F_{\sig }}{3} ( m_{\sig}^2 \mink_{\mu\nu} -   q_\mu q_\nu)  \;,
\end{equation}
 which defines $F_\sig$
 as the order parameter of spontaneous scale-symmetry breaking.
 This term contributes to the $D$-form factor, via a single dilaton exchange with the nucleon pair,
 parametrised by the on-shell interaction
 \begin{equation}
 \label{eq:GT}
 \de \Lageff  = g_{\sig NN} \, \sig \bar NN \;, \qquad  g_{\sig NN} = \frac{m_N}{F_\sig} \;.
 \end{equation}
 This result, derived  from \eqref{eq:Leff}, is independent of the Weyl-weight
and realises the Goldberger-Treiman mechanism \eqref{eq:Golberger} for the dilaton,  as
one readily obtains
\begin{equation}
\label{eq:rN}
D^N(q^2) = \frac{\res{\sig}{N}}{q^2} \;, \qquad
\res{\sig}{N} =   \frac{2}{3}  \bar u(p) u(p) \, F_\sig\,  g_{\sig NN}   = \frac{4}{3} m_N^2 \;,
\end{equation}
using $\bar u(p) u(p) = 2m_N$.

{After illustrating the mechanism, let us turn to an LO-estimate, including
   linear quark mass effects.  They have been determined in~\cite{Zwicky:2023fay}:\footnote{\label{foot:rewrite} Alternatively, one may write
 ${\gtwo}^\pi(q^2) = \frac{2}{3} \frac{m_\sig^2}{q^2- m_\sig^2}- \frac{1}{3}$, which is more natural from the  dispersive viewpoint.}}
\begin{equation}
\label{eq:Dold}
    \quad {\gtwo}^{N}(q^2) =  \frac{4}{3}  \frac{  \bar{m}_{N}^2 }{q^2-m_\sig^2}  \;, \quad
    {\gtwo}^\pi(q^2) =  \frac{2}{3}  \frac{  q^2   }{q^2-m_\sig^2} - 1 \;.
 \end{equation}
  The quantity $\bar{m}_{N}$ denotes the nucleon mass in the chiral limit 
   \begin{equation}
 {m}_{N} = \bar{m}_{N} + \de{m}_{N} \;,
 \end{equation}
 and $\de{m}_{N}$ the non-vanishing part due to  $m_{u,d,s} \neq 0$.
 In the nucleon case the linear quark corrections vanish since the residue is   of the form {
 \begin{equation}
 \label{eq:rNsig}
 \res{\sig}{N}  =  \frac{4}{3} m_N (\bar{m}_{N} - \de{m}_{N})  + \ORD(m_q^{3/2})  =  \frac{4}{3}( \bar{m}_N^2  - (\de{m}_{N})^2) + \ORD(m_q^{3/2}) \;,
 \end{equation}
 which relies on the determination of the fixed-point  anomalous dimension $\gast = 1$  of the quark mass \cite{Zwicky:2023bzk,Zwicky:2023krx}.
 The indicated corrections in \eqref{eq:rNsig} are radiative  and of the same form
as in baryon chiral perturbation theory \cite{Scherer:2012xha}, whereas in 
 the pion case they are of order {$\ORD(m_q^2 \ln m_q)$} as usual.

 {Regarding the pion}, it is worthwhile to point out that there is a {soft-pion theorem} constraining the LO trace to be  \cite{Donoghue:1991qv,Zwicky:2023fay}
\begin{equation}
\label{eq:low}
  \GaC^\pi(q^2)    =
  \begin{thincases}
2 m_\pi^2  + q^2  &   \quad  m_{\pi,\sig} \neq 0 \\[-0.2cm]
 0 &   \quad m_{\pi,\sig}  =  0
\end{thincases}  \;,
\end{equation}
and contracting \eqref{eq:GFF} this
 needs to match the expression
 \begin{equation}
  \label{eq:piontrace}
  \GaC^\pi(q^2)    = 2 m_\pi^2 A^\pi(q^2) - \frac{q^2}{2} ( A^\pi(q^2) + 3 D^\pi(q^2) )\;.
\end{equation}
Together with the model-independent     normalisation $A^\pi(0) = 1$, this  implies the
constraints
\begin{equation}
\label{eq:low2}
  \begin{thincases}
    D^\pi(0) = -1  \;, &  \quad  \quad  m_{\pi,\sig} \neq 0 \;\; \text{\cite{Donoghue:1991qv}}  \\[-0.2cm]
    D^\pi(0) = -\frac{1}{3} \;, \quad  \res{\sig}{\pi} = \ORD(q^2)  \;, & \quad   \quad m_{\pi,\sig}  =  0 \;\; \text{\cite{Zwicky:2023fay}}
\end{thincases}  \;,
\end{equation}
subject to higher order chiral corrections.
We emphasise that, for nonzero pion mass, the constraint holds independently of any fixed-point interpretation and is therefore directly relevant for the present work. In fact, Donoghue and
Leutwyler~\cite{Donoghue:1991qv} employed the soft theorem together with the $\sig$-meson dominance assumption to deduce this form. Consequently, the constraint $D^\pi(0) = -1$ provides a valuable handle for fixing the fit ansatz of the pion. In contrast, for the nucleon no analogous constraint exists, since constant or $q^2$-dependent shifts may arise from higher resonances or multi-particle contributions to the spectrum.

\section{The $\sig$-pole in the Euclidean }
\label{sec:dom}

Our primary fit ansatz consists of a $\sigma$-pole contribution together with a simple background term
\begin{equation}
D(q^2)  = \sig\text{-pole} + \text{background} \;.
\end{equation}
Describing the $\sig$-pole is a non-trivial task since
the $\sig$-meson is perhaps the most complicated and mysterious resonance of QCD \cite{Pelaez:2015qba}
and its pole is deep in the complex plane on the second sheet \cite{Caprini:2005zr}
\begin{equation}
\label{eq:pole}
\sqrt{s_\sig} \!=\!  m_\sig - \frac{i}{2} \Ga_\sig  =
(441^{+16}_{-8} \!-\!i272 ^{+9}_{-12.5})   \MeV \;.
\end{equation}
Our main point is that the details of the $\sig$-meson in the Minkowski domain are irrelevant in the deep Euclidean region, where a simple effective pole parametrisation proves sufficient.
In \SEC\ref{sec:ansatz}, we will revisit the issue and explain why alternative resonance parametrisations  are not well suited.

\subsection{The effect of a resonance in the deep Euclidean}
\label{sec:G}

Let us consider  a generic form factor
\begin{equation}
\form(q^2) = \matel{H(p')}{\Op}{H(p)} \;,
\end{equation}
where $\Op$ denotes the   operator,  transitioning from  $H(p)$ to $H(p')$,
and coupling  to  a stable particle $\party$  in the $q^2  = (p-p')^2$  channel.  
\com{We split the form factor into a part due to $\party$ and the rest 
$\form(q^2) = \form_\party(q^2) + \de \form(q^2)$. 
Since  $\form$ (and also $\form_\party$) are real for  $q^2 < m_\party^2$ (real analyticity), 
we may write the following dispersion representation 
\begin{equation}
\label{eq:Gdis}
\form_\party(q^2) = \int_{0}^\infty  \frac{ds \, \rho_\party(s)}{s-q^2-i0}  =
\frac{r_{0,\party}}{q^2-m^2_\party}   \;,
\end{equation}
 with spectral function 
$\rho_\party(s) = \tfrac{1}{\pi} \Ima G_\party(s)$.}
In this case, the effect of the pole across the complex plane is simple and fully controlled by the residue $r_{0,\party}$.
 Now suppose we turn on a parameter that renders the particle $\party$ unstable. The pole then moves to the second sheet, $\sqrt{s_\party} = m_\party - \frac{i}{2} \Ga_{\party}$, and one may define a  \emph{complex-valued}  residue $r_\party$ through
 \begin{equation}
 \label{eq:rC}
G_\party^{(\text{II})}(q^2)|_{q^2 \approx  s_\party}  = \frac{r_\party}{q^2-s_\party}  + \ORD(1)  \;,
 \end{equation}
where the superscript indicates the second-sheet continuation.   
\com{This complex pole, while interesting in its own right, does not provide an effective description of $\form_\party(q^2)$ in the Euclidean domain. This is evident from the fact that $\form_\party(q^2)$ is real in this region. From the dispersion relation one then finds
\begin{equation}
\label{eq:Euclid}
\form_\party(q^2) = \int_{0}^\infty \frac{ds , \rho_\party(s)}{s-q^2-i0} = \frac{\rE{\party}}{q^2- \mE{\party}^2} + \dots \;,
\end{equation}
where the dots denote terms that are suppressed for sufficiently Euclidean $q^2$ 
(see \APP\ref{app:multipole} for further details where it is phrased in terms of a   multipole-type expansion in momentum space).
 The effective residue is given by $\rE{\party} = - \int ds , \rho_\party(s)$. This quantity represents an average over the real spectral density of the particle and therefore cannot coincide with the complex-valued residue.}

Indeed, we anticipate
\begin{equation}
(i) \;\; \text{arg}( \sqrt{s_\sig}) \leftrightarrow \text{arg} (r_\party )
\;, \qquad (ii) \;\; \rE{\party} \neq |r_\party | \;, \qquad \com{ (iii) \;\; \rE{\party} \approx r_{\text{LO}} \;, }
\end{equation}
namely that the phase of the pole correlates with the phase of the residue, that the Euclidean residue can differ significantly from the modulus of the complex residue, \com{and that the effective residue is well approximated by the LO Lagrangian (provided the $\sig$-meson acts as a pseudo dilaton).} More generally, one would expect $\mE{\party} \approx |s_\party|$ and that
$|\rE{\party}/ r_\party | < 1$, with this effect becoming more pronounced as the width-to-mass ratio increases.
\com{ All of points (i), (ii) and (iii) are supported by the \LsM presented in \APP\ref{app:LsigM}, see in particular \TAB\ref{tab:resi}.}

\com{Point (ii) is crucial, since for the nucleon the complex residue is known and differs substantially in its absolute value from the dilaton residue; see \SEC\ref{sec:N} for a more detailed discussion. As point (iii) is central to our work, we would like to comment further. The LO result receives self-energy and vertex corrections, just as in the \LsM \eqref{eq:FNLO}. The effect of the self-energy is absorbed into the effective mass, while the vertex corrections induce a $q^2$ dependence that is difficult to compute reliably within the effective theory. The parametric dependence of the effective theory will enter the error estimates of the dilaton predictions discussed later on.  The leading corrections due to the quark mass are taken into account and 
missing corrections  again enter as a parametric uncertainty estimate.}

\subsection{Fit ansatz}
\label{sec:ansatz}

Since the lattice data are available in the Euclidean range $q^2 \in [-2,0] \GeV^2$, we adopt the following parametrisation for the $D$-form factor, motivated by the discussion in the previous section
\begin{equation}
\label{eq:D}
D(q^2)  =  \frac{\rE{\sig}}{q^2 - \mE{\sig}^2
} + \back(q^2)  \;.
\end{equation}
In fact, it has been noted in the literature that  $\sig$-effects in the Euclidean
domain are well approximated by a $\sig$-pole with pole-mass in the range $500$-$600 \MeV$
\cite{Donoghue:2006rg} (see also \cite{RuizArriola:2010fj,Broniowski:2024oyk,Broniowski:2025ctl}), further supported by the empirical success of one-boson-exchange models (e.g.~\cite{CalleCordon:2008eu,Wu:2023uva}).
The fact that the up and down quark masses are expected to enlarge the real
part of the complex pole might well be compensated by the reduction of the imaginary part because of reduced phase space.
We therefore ignore this effect and adopt $\mE{\sig} = 550(50) \MeV$ as our central value.
The parameter $\rE{\sig}$ serves as our primary fit variable, to be tested against the prediction of  dilaton effective theory.

The background contribution must of course be nonzero, as it accounts for higher states in the spectrum. In a spontaneously broken conformal theory, the dilaton (identified here with the $\sig$-meson) is the only state that couples in the scalar channel, as follows from the textbook derivation of Goldstone’s theorem \cite{Weinberg:1996kr}. By contrast, the situation is less clear in a theory flowing to an infrared fixed point with spontaneous breaking of scale invariance. In such a scenario, suppression of other states near the fixed point appears plausible, since in the effective theory they would be loop-suppressed, unlike for the dilaton.  \comb{We may gain insight by employing the dispersive techniques of \cite{Cao:2024zlf,Cao:2025dkv}, which are based on unitarity cuts and a two-channel Omn\`es solution, to infer the $D$-form factor (ultimately from experimental data and Roy–Steiner equations). For this purpose it is convenient to introduce a normalised trace of the energy-momentum tensor  $\hat{\Theta}^N(q^2)\equiv \Theta^N(q^2)/(2m_N^2)$, $\hat{\Theta}^N(0)=1$, and to consider the unsubtracted dispersion representation
\begin{equation}
\label{eq:trace}
\Theta^N(q^2) = \frac{1}{\pi} \int_{4 m_\pi^2}^\infty \frac{ds \, \Ima \Theta^N(s) }{(s-q^2-i0)} \;.
\end{equation}}

\comb{In order to make optimal use of data, the $\pi\pi$ and $K\bar K$ coupled-channel system is analysed in \cite{Cao:2024zlf,Cao:2025dkv}, together with an effective pole accounting for higher states, a strategy with a long-standing tradition, for instance in the description of electromagnetic form factors. To assess the $\sig$-contribution, however, we restrict ourselves to the $\pi\pi$ channel, since the second resonance, $f_0(980)$, is known to predominantly couple to $K\bar K$. We therefore approximate  the dispersive integral
\begin{equation}
\Theta^N(q^2) \approx
\frac{1}{\pi} \int_{4 m_\pi^2}^{4 m_K^2} \frac{ds \, \Ima \Theta^N(s) }{(s-q^2-i0)} + \frac{c_{\Theta}}{1-q^2/m_S^2} \;,
\end{equation}
by cutting at the $K\bar K$ threshold, thus avoiding the coupled-channel region, and adding an effective pole that accounts for $f_0(980),f_0(1370)$ and further states. The single-channel imaginary part is given by \cite{Cao:2024zlf,Cao:2025dkv}
\begin{align}
\label{eq:ima}
\Ima \Theta^N(s)=-\frac{3\rho_\pi}{2s \rho_N^2}\left(f^0_+(s)\right)^*\Theta^\pi(s) \;, \quad
\rho_x(s) \equiv \sqrt{1- 4 m_x^2/s} \;,
\end{align}
where $\Theta^\pi(s)$ follows from the Omn\`es-  and $f^0_+(s)$ from the $\pi N$ Roy–Steiner-solution. 
These quantities play the roles of $F_\sig$ and $g_{\sig NN}$ in \eqref{eq:rN}.}
     
\comb{ The normalisation of the effective pole is fixed by the condition $\Theta^N(0)=1$, yielding 
 $c_{\Theta}\approx -0.19(12)$ with uncertainties originating from the input to \EQ\eqref{eq:ima}.\footnote{We are indebted to the authors of \cite{Cao:2024zlf} for sharing their data with us.} 
 The fact that the $\sig$-meson contribution alone, within the one-channel approximation, saturates the sum rule reasonably well is remarkable and reminiscent of vector-meson dominance in the spin-$1$ channel, which itself still lacks a deeper explanation. It is therefore worth noting that, in the spontaneously broken scale invariance scenario, $\sig$-dominance arises naturally, at least in the trace of the energy-momentum tensor.}
 
\comb{Since we are fitting the $D$-form factor, its additional contributions must also be taken into account. These follow straightforwardly from the definition \eqref{eq:gffdefinition}
\begin{equation}
D^N(q^2) = \frac{4 m_N^2}{3 q^2}( A^N(q^2) -\hat{ \Theta}^N(q^2) ) - \frac{1}{3}( A^N(q^2) - 2 J^N(q^2)) \;.
\end{equation}
One observes that, in addition to the spin-$0$ channel, this expression involves the form factors $A$ and $J$, which are of spin-$2$ type, see for example \cite{Broniowski:2025ctl}. The second term vanishes at $q^2=0$ and remains small throughout, whereas the $A$-form factor in the first term is sizeable, with $f_2(1270)$ being the first resonance in the spectrum. }

\comb{We therefore conclude that, for the $D$-form factor, there is empirical evidence that the $\sig$-meson is highly dominant in its spin-$0$ component, 
while the leading spin-$2$ resonance, $f_2(1270)$, appears at significantly higher mass. Moreover, we have verified that the $A$ (and $J$) form factors are very well described by a quadratic polynomial in $q^2$, which 
altogether motivates the parametrisation
\begin{equation}
\label{eq:back}
\back(q^2) = b + b' q^2 + b'' q^4 + \frac{r_{\text{eff}}}{q^2 - m_{\text{eff}}^2} \;,
\end{equation}
with $m_{\text{eff}} = 1.2 \GeV$  since our analysis indicates that $f_2(1270)$ is more prominent than $f_0(980)$. We note that \eqref{eq:back} constitutes an over-parametrisation and would therefore lead to inflated uncertainties. Consequently, we restrict our fits to reduced parameter sets, typically involving three parameters including the $\sig$ residue, and use the various fit combinations to assess robustness.}

\comb{Finally, we emphasise that this background does not reflect a systematic effective-theory expansion in $q^2/(4 \pi F_{\pi,\sig})^2$ which would have a radius of convergence well below $2\GeV^2$. 
Instead, the background, which also includes a constant term, effectively parametrises a sum of the most relevant  higher resonances and multi-hadron states.}

Before presenting the fits, we digress to comment in more technical language
 why certain alternative parametrisations  are not employed. 
\begin{itemize}
\item [(i)]
Are there viable alternatives to the Euclidean pole parametrisation?
A first difficulty arises from the fact that, although the pole position itself is uniquely defined, the behaviour in its vicinity is not as it depends sensitively on the production mechanism or process under consideration. This issue is particularly relevant for the $\sig$-meson, which is both broad~\eqref{eq:pole} and located close to the left-hand cuts of $\pi\pi$-scattering. These features explain why the $\sig$ phase shift does not pass rapidly through $180^\circ$; see, for example,
\FIG 2 in \cite{Pelaez:2015qba} or the discussion in~\cite{Zheng:2003cr}.
 Through Watson’s theorem, this behaviour directly carries over to the form-factor case, since the same phase enters  both $\pi\pi$-scattering and the form factor (via the Omn\`es representation \cite{Omnes:1958hv}),
see for instance \cite{Moussallam:2011zg,Hoferichter:2012wf}. This property alone rules out many commonly used parametrisations, including all Breit–Wigner forms and  Flatt\'e-type (or ``sill")  models \cite{Giacosa:2021mbz}. While the sill model performs reasonably well for the moderately broad $a_1(1260)$-meson, it fails for the much broader $\kappa/K_0^*(700)$, which in fact shares important characteristics with the $\sig$~\cite{Giacosa:2021mbz}.
Other approaches, such as the K-matrix employed by the HadSpec collaboration at $m_\pi \approx 239 \MeV$~\cite{Rodas:2023nec}, or the single-resonance $S$-matrix solution of \cite{Zheng:2003cr}, are not suitable either, since they are designed for the $S$-matrix rather than for form factors. In scattering amplitudes, left-hand cuts are naturally included in these frameworks, which are absent in  the form-factor case.  A dispersive analysis using the Omn\`es representation \cite{Cao:2024zlf,Cao:2025dkv}, is closer to extracting
the complex-valued residue which is not what we are aiming at.

\item [(ii)] It has been argued that in the deep Euclidean limit $q^2 \to -\infty$, the form factors scale as $D^\pi \propto 1/q^2$ and $D^N \propto 1/q^6$, based on light-cone sum rule computations~\cite{Tong:2021ctu,Tong:2022zax}; see also~\cite{Liu:2024vkj} for an instanton-based
approach to the pion case. In the absence of further input, such asymptotic constraints can provide useful guidance for parametrisations~\cite{Broniowski:2024oyk,Broniowski:2025ctl}. However, we choose not to impose them here for several reasons.
First, it is not evident that $q^2 = -2 \GeV^2$ should already be regarded as the deep Euclidean regime, and most of our relevant fit points are even less Euclidean \comb{(see also the remarks in \REF\cite{Leutwyler:2002hm}).} Second, the light-cone sum rule estimates rely heavily on the endpoint region of the corresponding distribution amplitudes, which are generally unknown and often assumed to take the asymptotic form. This assumption has been challenged by high-$q^2$ measurements of $\eta(\pi)\ga\ga^*$ transitions; see, for example,~\cite{Agaev:2014wna} and references therein.
\end{itemize}

\section{Numerics and Fits to Lattice Data at $m_\pi \approx 170 \MeV$ }
\label{sec:fits}

In this section, we fit our ansatz~\eqref{eq:D} to lattice data at $m_\pi \approx 170 \MeV $~\cite{Hackett:2023rif,Hackett:2023nkr}, close to the physical pion mass of
$m_\pi \approx 140 \MeV$. Our goal is to test whether the $\sigma$-meson ansatz naturally reproduces the $\sig$-residue predicted by the dilaton interpretation \EQ\eqref{eq:Dold}.
Attempting the converse, deriving the dilaton picture directly from the data, proves
too challenging for the pion and only marginally feasible for the nucleon.

\subsection{The nucleon gravitational form factor}
\label{sec:N}

The concrete  fit ansatz for the nucleon gravitational form factor is
\begin{equation}
\label{eq:Nansatz}
D^N(q^2) = \frac{\rE{\sig}^N}{q^2 - \mE{\sig}^2}+
b + b^{'}q^2 + b^{''}q^4 +
\frac{ \reff}{q^2 - \meff^2} \;,
\end{equation}
where  $\meff = 1.2 \GeV$   as argued around \EQ\eqref{eq:back}. 
 In our main fit we set $\reff = 0$, while alternative combinations are used to test the robustness 
of the residue with respect to background variations. 
 Since our primary goal is to compare the fit result with the dilaton prediction
$\res{\sig}{N} = \frac{4}{3} \bar{m}_N^2$ \eqref{eq:Dold}, we need to estimate its central value and associated uncertainty. Here, $\bar{m}_N$ denotes the nucleon mass in the limit of vanishing $m_{u,d,s}$, for which we adopt the value $\bar{m}_N = 826 \MeV$ from the recent pedagogical introduction \cite{Hoferichter:2025ubp}.\footnote{The uncertainty in $\bar{m}_N = 826 \MeV$ is small but difficult to assess because of the longstanding tension between phenomenology and lattice results regarding light quark contributions, see~\cite{FlavourLatticeAveragingGroupFLAG:2024oxs,Hoferichter:2025ubp} for references. A possible explanation could be excited state contamination \cite{Gupta:2021ahb}.} This corresponds to a $13\%$ reduction from the physical nucleon mass, $m_N \approx 938 \MeV$,
with a significant part arising from the nucleon's strangeness content.
As discussed earlier, the leading parametric corrections to the nucleon residue,
beyond known $\ORD(m_q)$,
are   radiative corrections  order of $\ORD(m_q^{3/2})$ \eqref{eq:rNsig}. {In order to estimate them we will take the nucleon mass
corrections as a guidance. At $m_\pi \approx 170\MeV$ the nucleon mass is expected to be around
$970\MeV$ as can be deduced from the mass decomposition \cite{Hoferichter:2025ubp}
 or lattice plots for instance
\cite{Owa:2023tbk}. We will conservatively assign $50\%$ of this to the unknown  $\ORD(m_q^{3/2})$ and higher corrections, which amounts to  a  $72\MeV$-nucleon mass   and $0.08\GeV^2$-residue uncertainty, respectively. 
\com{Let us return to the $q^2$-dependent vertex corrections mentioned at the end of \SEC\ref{sec:G}, which we estimate via the parametric dependence in the effective theory given by $q^2/(4 \pi F_\sig)^2$. From $F_\sig^2 \propto N_f$ and since $F_\pi = F_\sig$ in the $N_f = 2$ linear $\sig$-model, we take $F_\sig^2 = 3/2, F_\pi^2$. The pion decay constant at $m_\pi \approx 170 \MeV$ is $F_\pi \approx 100\MeV$, as inferred from the plots in \cite{Brandt:2013dua}. Since $q^2/(q^2 - \mE{\sig}^2) = \mE{\sig}^2/(q^2 - \mE{\sig}^2) + 1$ and the constant term can be absorbed into the background, this implies a relative uncertainty of $\mE{\sig}^2/(4 \pi F_\sig)^2 \approx 0.13$. Adding both uncertainties in quadrature  yields the estimate
\begin{equation}
\label{eq:rNtheory}
\res{\sig}{N}|_{\text{dilaton}} = 0.91(14) \GeV^2 \;.
\end{equation}
}
For the fit, we minimise the $\chi^2$
\begin{equation}
\label{eq:chi2}
\chi^2 = \sum_{i,j=1}^{N_{\mathrm{data}}}
(D^N_{\text{data}} - D^N_{\text{model}})_i \,
( \text{Cov}^{-1})_{ij}  \, (D^N_{\text{data}} - D^N_{\text{model}})_j \;, \qquad
\hat{\chi}^2 \equiv \frac{\chi^2}{N_\mathrm{dof} }\;,
\end{equation}
using all  33 available data points \cite{Hackett:2023rif}. 
The effective number of degrees of freedom is $N_\mathrm{dof} =30$ since there are three fit parameters.
A fit using the main parametrisation is displayed in the left figure of \FIG\ref{fig:170mevfits}, where it is compared to the data of \REF\cite{Hackett:2023rif}. Table~\ref{tab:N170} shows the corresponding fit parameters for a range of effective $\sig$-masses, and the uncertainties correspond to the standard deviations as encoded in the covariance matrix.
The covariance matrix of the fitted parameters is estimated from the inverse $\chi^2$-Hessian, evaluated at the $\chi^2$-minimum.

We identify two primary sources of uncertainty: the data itself and the choice of $\mE{\sig}$. The data uncertainty is propagated to the fit parameters through the fit procedure, while the dependence on $\mE{\sig}$ can be assessed by varying its value within $\mE{\sig}= 550(50)\MeV$.
The background parameters $b$ and $b'$ are treated as nuisance parameters in the fit, and their variation is accounted for in the quoted uncertainty on $\rE{\sig}^N$.

The final result is
\begin{equation}
\label{eq:rNfit}
\rE{\sig}^N = 1.13(26)(20) \GeV^2 \;,
\end{equation}
which is consistent with the dilaton prediction given in \eqref{eq:rNtheory}.

\begin{table}[h]
\centering
\begin{tabular}{  c  | c  |  c  | c | c}
$\mE{\sig}\, [\MeV]$ & $\rE{\sig}^N \, [\GeV^2] $  & $ b$ & $b^{'}$ & $\chi^2/N_\mathrm{dof}$ \\  \hline
450   & 0.83(19)      & 0.44(30) & 0.10(13) & 0.55 \\
500   & 0.97(22)      & 0.56(32) & 0.13(14) & 0.55 \\
  \rowcolor{gray!15} 550   & 1.13(26)      & 0.69(35) & 0.17(14) & 0.55 \\
600   & 1.32(30)      & 0.84(38) & 0.21(15) & 0.56 \\
650   & 1.54(35)      & 0.99(42) & 0.25(16) & 0.56 \\
\end{tabular}
\caption{\small
  Model parameters for the nucleon $D$-form factor resulting from fits using our main background parametrisation, and for various values of the pole mass $\mE{\sig},$ to the 33 lattice data points at $m_\pi \approx 170\MeV$~\cite{Hackett:2023rif}.
   The uncertainties correspond to the standard deviations encoded in the covariance matrix. The correlations between parameters
  are  $(\rho_{rb}, \, \rho_{rb'}, \, \rho_{bb'}) = (0.97, \, 0.91, \, 0.98)$ for $\mE{\sig} = 550 \MeV$.
   LO dilaton effective theory predicts $\res{\sig}{N}=0.91(14)\GeV^2$~\eqref{eq:rNtheory}.
}
\label{tab:N170}
\end{table}

Let us turn to testing the robustness of the fit.
Allowing all three background parameters to vary simultaneously leads to large marginalised uncertainties which would make a comparison to the dilaton prediction largely inconclusive. On the other hand, omitting the background entirely is theoretically disfavoured, as discussed above.
Insisting on  no background, the fit yields $\rE{\sig} = 0.476(34)$,  with a $\chi^2$-value  that is nearly double that of the main parametrisation, which indeed rules out this ansatz on its own. More reasonable combinations, fitted with a fixed $\mE{\sig} = 550\MeV$, are
\begin{itemize}
\item [a)] $\{ \rE{\sig}, b \}^N
= \{ \, 0.86(11)\, , \, 0.291(77) \}$ and $\hat{\chi}^2 = 0.58$
\item [b)] $\{ \rE{\sig}, b, \reff \}^N
= \{ \, 1.40(46)\, , \, -0.13(35)\, , \, -2.1(1.7) \}$ and $\hat{\chi}^2 = 0.55$
\item [c)] $\{ \rE{\sig}, \reff \}^N
= \{ \, 1.25(20)\, , \, -1.49(38) \}$ and $\hat{\chi}^2 = 0.54$
\item [d)] $\{ \rE{\sig}, b', \reff \}^N
= \{ \, 1.36(36)\, , \, 0.026(74)\, , \, -1.78(89) \}$ and $\hat{\chi}^2 = 0.55$
\item [e)] $\{ \rE{\sig}, b, b', b'' \}^N
= \{ \, 1.37(57)\, , \, 1.2(1.1)\, , \, 0.026(74)\, , \, 0.12(26) \}$ and $\hat{\chi}^2 = 0.57$
\end{itemize}
Firstly, we note that all of these fits are compatible with the main result \eqref{eq:rNfit}. Secondly, they also remain consistent with the theoretical prediction \eqref{eq:rNtheory}, particularly once the spread in the effective mass is taken into account. Fits c) and d) provide a more direct test of the $\sig$-meson dominance hypothesis, as the $\reff$-residue could potentially account for the curvature in the data. We observe that, while $\reff$ is non-negligible, $\rE{\sig}^N$ continues to agree with the theoretical expectation. \comb{We further note that reducing the fit interval to $[0,-1.5]\GeV^2$ and $[0,-1]\GeV^2$ for the linear background leads to $\{ \rE{\sig}^N , \hat{\chi}^2 \} = \{ 1.19(34) , 0.6 \}$ and $\{ \rE{\sig}^N , \hat{\chi}^2 \} = \{ 1.34(72) , 0.46 \}$, respectively, which further hints at the robustness of the result.}
In conclusion, there is good agreement within uncertainties with the dilaton interpretation, and this last observation suggests that the reverse scenario, establishing
$\sig$-dominance,  is not out of sight.
\begin{figure}[h]
  \centering
  \includegraphics[width=0.49\textwidth]{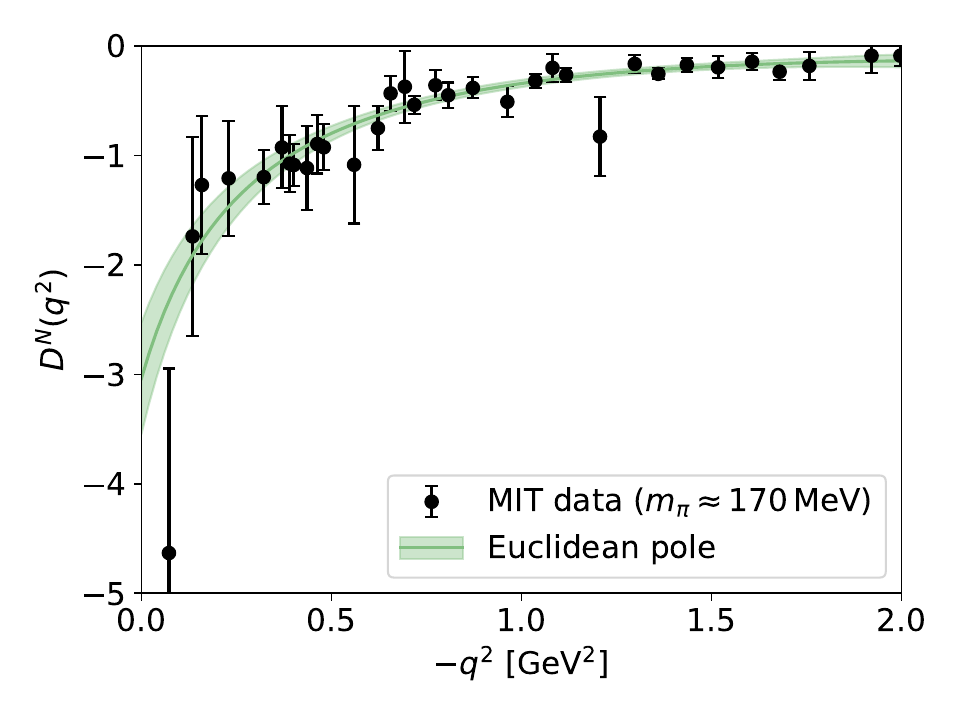}
  \includegraphics[width=0.49\textwidth]{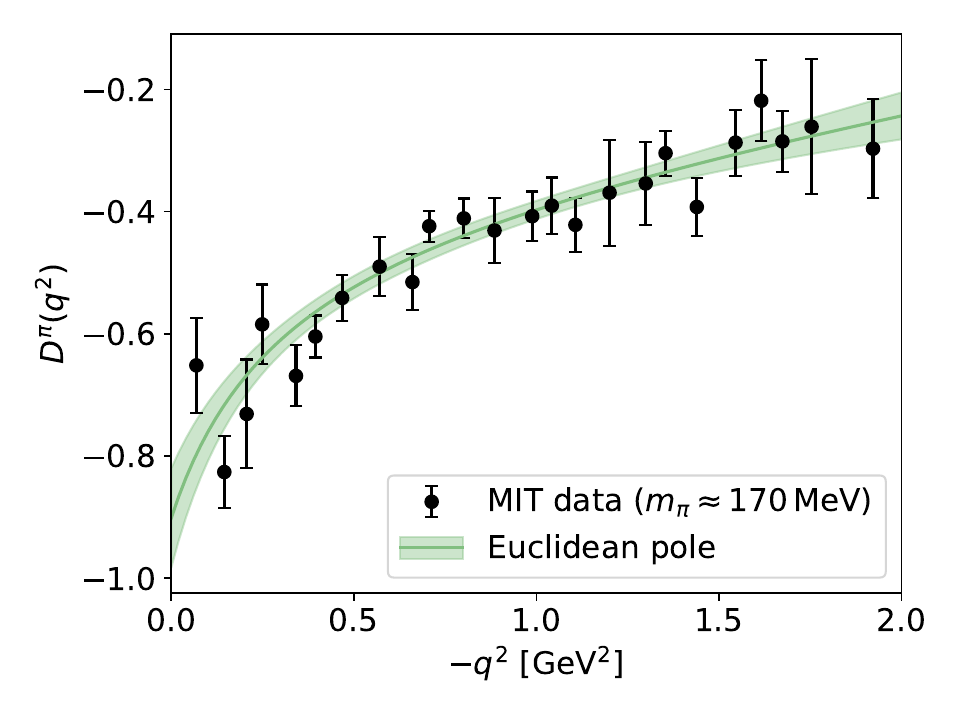}
  \caption{\small The $D$-form factors are fitted using a Euclidean pole parametrisation
  \eqref{eq:Nansatz} and  \eqref{eq:pionfunc}
  with {$\reff=0$ and} $\mE{\sig}=550\MeV$. The fits are compared to the lattice data, shown in black, for the nucleon \cite{Hackett:2023rif} (left) and for the pion \cite{Hackett:2023nkr} (right). The dark  curve indicates the central fit, while the shaded band represents the $68\%$ confidence interval.}
    \label{fig:170mevfits}
\end{figure}

Now, the complex-valued residue has been determined using Roy-Steiner equations
 \cite{Hoferichter:2023mgy}. Their result and our  fit to the Euclidean residue  differ, also
 in their absolute values
\begin{equation}
\label{eq:r2}
\res{\sig}{N}|_{\mbox{\cite{Hoferichter:2023mgy}}} = (0.90(28)- 2.78(20)i) \, \left[ \frac{4}{3}  m_N^2 \right]
\;, \quad   \rE{\sig}^N|_{\eqref{eq:rNfit}}  \approx 1.04    \,   \left[ \frac{4}{3}  m_N^2 \right]  \;.
\end{equation}
 Following the discussion in  \SEC\ref{sec:G}, and particularly in our explorations within the \LsM in \APP\ref{app:broad},  this should not come as a surprise.
The complex-valued residue at the pole and the effective residue in the Euclidean region
are simply not the same quantity.
Hence, the two values in \eqref{eq:r2} might well originate from the same underlying theory.

\subsection{The pion gravitational form factor }

For the pion gravitational form factor we adopt the ansatz
\begin{equation}
\label{eq:pionfunc}
D^\pi(q^2) = \frac{q^2 \, \rE{\sig}^\pi}{q^2 - \mE{\sig}^2} - 1 + \hat{b} + b^{'}q^2 + b^{''}q^4 +  \frac{\reff}{q^2 - \meff^2} \;,
\end{equation}
where both the prefactor $q^2$ in the numerator and the $-1$ term are dictated by the soft-pion theorem \eqref{eq:low2}. This provides a robust constraint, subject only to chiral corrections, implying that
\begin{equation}
\label{eq:soft-fit}
\text{soft-pion theorem} \quad \Rightarrow \quad \hat{b} - \frac{\reff}{\meff^2} = 0 \pm 0.1 \;,
\end{equation}
which the fits satisfy. An alternative would be to impose the constraint along with next-to-leading order (NLO) computation in chiral perturbation theory  \cite{Donoghue:1991qv}.

As in the nucleon case, we set $\reff=0$ in the main fit. The theoretical prediction for the residue is $\res{\sig}{\pi} = 2/3$  
\eqref{eq:Dold}\com{For the uncertainties, we include the $13\%$ $q^2$-dependent corrections, as in the nucleon case. Pion mass corrections, $\ORD( [m_\pi^2 \ln m_\pi]/( 4\pi F_\pi)^2)$, are about four times smaller, so that adding the contributions in quadrature leads to
\begin{equation}
\label{eq:rPItheory}
\res{\sig}{\pi}|_{\text{dilaton}} = \frac{2}{3} \pm 0.1 \;.
\end{equation}
}
\begin{table}[h]
 \centering
\begin{tabular}{  c  | c  |  c  | c | c }
$\mE{\sig}\, [\MeV]$ & $\rE{\sig}^\pi $  & $ \hat{b}$ & $b^{'}$ & $\chi^2/N_\mathrm{dof}$ \\  \hline
450   & 0.48(16)      & 0.070(94) & -0.130(45) & 1.23 \\
500   & 0.50(16)      & 0.085(87) & -0.116(48) & 1.20 \\
 \rowcolor{gray!15} 550   & 0.53(16)      & 0.098(81) & -0.101(51)  & 1.18 \\
600   & 0.56(17)      & 0.109(77) & -0.086(55) & 1.16 \\
650   & 0.59(18)      & 0.119(73) & -0.070(58) & 1.14 \\
\end{tabular}
\caption{\small
  Same as \TAB\ref{tab:N170}, but now for the pion data~\cite{Hackett:2023nkr} fitted using the parametrisation \eqref{eq:pionfunc}.
  LO dilaton effective theory predicts $\res{\sig}{\pi}=2/3$.
  The correlations between parameters for $\mE{\sig} = 550 \MeV$ are $(\rho_{r\hat{b}}, \, \rho_{rb'}, \, \rho_{\hat{b} b'}) = (-0.94, \, 0.86, \, -0.65)$.
}
\label{tab:pion170}
\end{table}

We apply the same fitting procedure as for the nucleon using all 24 data points of \REF\cite{Hackett:2023nkr}, with $N_\mathrm{dof}=21$ since there are three fit parameters.
The main fits are shown in \FIG\ref{fig:170mevfits} and reported in \TAB\ref{tab:pion170} for a range of effective $\sig$-masses, with uncertainties represented as in \TAB\ref{tab:N170}.
The dependence on the effective mass is much smaller in this case, and our main fit result is
\begin{equation}
\label{eq:rPIfit}
\rE{\sig}^\pi = 0.53(16)(3) \;.
\end{equation}

As for the nucleon, we test robustness by considering different background parametrisations.
Fitting with no background is better motivated here, since the constant term is fixed by the soft-pion theorem.
Indeed, setting $\hat{b} = b' = \reff = 0$ yields
$\rE{\sig}^\pi = 0.8(2)$ with $\hat{\chi}^2 = 1.24$, not far from the fit-result in the  table.
The same background combinations as in the nucleon case give
\begin{itemize}
\item [a)] $\{ \rE{\sig}, \hat{b} \}^\pi
= \{ \, 0.80(8)\, , \, -0.006(62) \}$ and $\hat{\chi}^2 = 1.30$
\item [b)] $\{ \rE{\sig}, \hat{b}, \reff \}^\pi
= \{ \, 0.24(27)\, , \, 0.81(38)\, , \, 0.96(43) \}$ and $\hat{\chi}^2 = 1.13$
\item [c)] $\{ \rE{\sig}, \reff \}^\pi
= \{ \, 0.81(4)\, , \, 0.03(7) \}$ and $\hat{\chi}^2 = 1.29$ 
\item [d)] $\{ \rE{\sig}, b', \reff \}^\pi
= \{ \, 0.57(14)\, , \, -0.11(6)\, , \, -0.13(11) \}$ and $\hat{\chi}^2 = 1.19$
\item [e)]  $\{ \rE{\sig},\hat{b},b',b'' \}^\pi = \{ \,  0.23(35)\, , \, 0.16(11),\, -0.37(29)\, ,\;
-0.10(11) \}$ and  $\hat{\chi}^2 = 1.19$
\end{itemize}
 First, it is reassuring, and a good indication of the quality of the data, that all fits satisfy the model-independent soft-pion theorem constraint \eqref{eq:soft-fit}.
Second, with the exception of case b), all fits are compatible with our main result, especially once the spread in $\mE{\sig}$ is taken into account.
Case c) adds little compared to the no-background fit, since its single parameter is effectively constrained to be small by the soft-pion condition.

The situation is different for case b). Unlike for the nucleon, the extracted residue is \comb{marginally} compatible with the dilaton prediction.
This indicates that the converse approach, establishing $\sig$-dominance directly, is not feasible for the pion.
Rewriting the result as per footnote~\ref{foot:rewrite}, one finds
a residue  of {$\res{\sig}{\pi} \to \frac{2}{3} m_\sig^2 \approx 0.2 \GeV^2$ (using $m_\sig \to \mE{\sig}$)}, far smaller than both the nucleon residue \eqref{eq:rNtheory} and the potential $f_0(980)$ contribution in the pion case.
This outcome is therefore not surprising and also explains why the fits are less sensitive
to the effective mass $\mE{\sig}$ than in the nucleon case.
\comb{As for the nucleon, we reduce the fit interval to $[0,-1.5]\GeV^2$ and $[0,-1]\GeV^2$ for the linear background. For these choices we obtain $\{ \rE{\sig}^\pi , \hat{\chi}^2 \} = \{ 0.56(19) , 1.45 \}$ and $\{ \rE{\sig}^\pi , \hat{\chi}^2 \} = \{ 0.32(28) , 0.46 \}$, respectively, which is reasonably stable
 but at the same time confirms our finding that the pion case is slightly less robust than the nucleon case.}

In conclusion, we find that the pion fits are overall consistent with the dilaton interpretation within uncertainties, although, unlike in the nucleon case, the $\sig$-dominance hypothesis cannot be inferred.

\subsection{A quick note on the $D$-form factor in the infrared ---  the $D$-term}
\label{sec:note}

As previously discussed, conserved-current form factors are often connected with simple physical interpretations in the infrared.
For instance, the gravitational form factors satisfy $A(0)=1$ and $J(0)=\frac{1}{2}$, reflecting their association with energy and angular momentum.
By contrast, an interpretation of the $D$-form factor has long been elusive; for reviews, see \cite{Polyakov:2018zvc,Lorce:2025oot}.
Particular emphasis has been given to $D \equiv D(0)$, commonly referred to as the $D$-term (or Druck-term).
Viewing the $\sig$-meson as a dilaton, combined with phenomenological fits, provides a new perspective on this longstanding problem.

If the $\sig$ were to become massless in the chiral limit, the nucleon $D$-form factor develops a pole \eqref{eq:Dold},
\begin{equation}
D^N(q^2) = \frac{4}{3} \frac{m_N^2}{q^2} + \ORD(1) \;,
\end{equation}
as emphasised earlier.
While this scenario is generally not considered likely, there is also no definitive evidence ruling it out, and recent $N_f = 4$ quark-mass-degenerate lattice simulations do certainly not exclude
a massless $\sig$~\cite{LatKMI:2025kti}.\footnote{See also \cite{Goertz:2024dnz} for functional methods or   \cite{Elander:2025fpk} for 
holographic approaches.}

In the real world, where the $\sig$ acquires a nonzero mass, at least  due to light quarks, the pole disappears and the $D$-term remains finite.
In the Breit frame, the $D$-term can be expressed in terms of the pressure and shear forces  \cite{Polyakov:2002yz}.
When supplemented with the assumption of mechanical stability \cite{Perevalova:2016dln}, this  formulation implies $D< 0$ (or even $D < -0.20(2)$~\cite{Gegelia:2021wnj}),
a feature observed in many hadronic systems though not, for example, in hydrogen \cite{Ji:2022exr,Czarnecki:2023yqd,Freese:2024rkr};  see  \cite{Lorce:2025oot} for a thorough review.
 In our approach we obtain \eqref{eq:D},
  \begin{equation}
  \label{eq:Ddil}
 D \equiv D^N(0) = -  \frac{\rE{\sig}^N}{\mE{\sig}^2} + \back(0)  \;,
  \end{equation}
 and with
 \begin{equation}
 \rE{\sig}^N  \approx \res{\sig}{N} = \frac{4}{3} \bar{m}_N^2  > 0 \;,
 \end{equation}
 one infers that the $\sig$-contribution to the nucleon $D$-term is necessarily negative.
  Our fit yields $D^N(0)|_{\text{fit} }= -3.0(5)(3)$ (with the second uncertainty
  corresponding to the $\mE{\sig}$-variation) which breaks up into
  a $\sig$- and a background-contribution as $D^N(0)|_{\sig,\text{fit} }= -3.74(86)(66)$
  and $b(0)_\text{fit}=0.68(35)$ (cf. \TAB\ref{tab:N170}),
  and our dilaton prediction for the $\sig$-contribution is  $D^N(0)|_{\sig} = -3.01(39)$ \eqref{eq:rNtheory}.
This leads to the conclusion that, provided the $\sig$-meson dominates over the background, the nucleon $D$-term is negative.

Other values obtained from the lattice data include the $z$-expansion fit $D = -3.35(58)$
in the original paper \cite{Hackett:2023rif} and $D = -3.0(4)$  from a
 constrained tripole-fit \cite{Broniowski:2025ctl}. Furthermore, a dispersive analysis at physical pions mass was used to obtain a value of $D =
 -3.38^{+0.34}_{ -0.35}$ \cite{Cao:2024zlf}.
The close agreement among these results is reassuring.
However, our main point is not the value but that the $\sig$-contribution is necessarily negative
in the dilaton picture, and if dominant over the background, it implies a negative $D$-term.
It is in this respect that our work differs from others.

The $D$-term is also sought after since it enters the radii associated with the
energy-momentum tensor. The mean square radius in the Breit-frame and the mass radius read~\cite{Polyakov:2018zvc}
\begin{equation}
\vev{r^2_{\Theta}} = 6 A^{'}(0) - \frac{9}{2 m_N^2} D \;, \quad
\vev{r^2_{\text{mass}}} = 6 A^{'}(0) - \frac{3}{2 m_N^2} D  \;,
\end{equation}
respectively.  From \eqref{eq:Ddil} we infer that the $\sig$-contribution is inversely proportional
to the effective $\sig$-mass
\begin{equation}
 \vev{r^2_{\Theta}}|_{\sig} = 6( A'(0) + \frac{1}{\mE{\sig}^2} \frac{\bar{m}_N^2}{m_N^2} - \frac{3}{4} \frac{b(0)}{m_N^2} ) \;.
\end{equation}
 Thus the mass radius gets larger as the effective $\sig$-mass
decreases. This is intuitive and analogous to the pion charge radius, although there the dependence
is only logarithmic  on the pion mass \cite{Donoghue:1992dd}.

What about the $D$-term for other hadrons?
For non-Goldstone states the same pattern applies, see \eqref{eq:Golberger}.
In particular, we find similar behaviour for the gluonic gravitational form factors of the
$\rho$-meson   and the  $\De$-baryon \cite{Stegeman:2025tdl},
based on lattice data at $m_\pi \approx 450 \MeV$ \cite{Pefkou:2021fni}.
Let us turn to the Goldstone case.
For the pion, the soft-pion theorem requires $D^\pi(0) = -1$ \cite{Donoghue:1991qv}.
However, in the presence of a massless $\sig$, this relation is modified to $D^\pi(0) = -\tfrac{1}{3}$ \cite{Zwicky:2023fay},
as follows directly from \eqref{eq:Dold}, since the dilaton pole prevents a naive application of the soft-pion theorem.\footnote{For a massive $\sig$, one gets
$D^\sig(0) = -\frac{3+2 \De_{\cal O}}{3}$  where  ${\cal O}$ is the operator generating the mass.
 If  ${\cal O} = m_q \bar qq$ then $\De_{\cal O} = 2$ and $D^\sig(0) = - \frac{7}{3}$.
For a massless $\sig$, one has   $D^\sig(0) = -\frac{1}{3}$ as for the pion \cite{Zwicky:2023fay}.}

\section{Conclusion and Discussion}
\label{sec:conc}

In this work, we have fitted the   pion and the nucleon gravitational form factors (from lattice QCD data \cite{Hackett:2023rif,Hackett:2023nkr}) to a parametrisation  \eqref{eq:D}
where the $\sig$-meson plays a central role.
The main quantity of interest is the residue $\rE{\sig}$
of the effective $\sig$-pole  which we
compare to the  dilaton effective theory predictions \eqref{eq:Dold}.
The fit ansatz was motivated  theoretically and tested specifically in the \LsM
 which served as a toy model, allowing us to understand
 the difference between the complex-valued residue and the effective  Euclidean residue.

For the nucleon, we find that the fitted $\sig$-residue is compatible with  the dilaton effective theory
prediction within uncertainties. Specifically, \EQs\eqref{eq:rNfit} and \eqref{eq:rNtheory}  are
\begin{equation}
\label{eq:N}
\rE{\sig}^N = 1.13(26)(20) \GeV^2  \;, \qquad \res{\sig}{N}|_{\text{dilaton}}  = 0.91(14) \GeV^2  \;,
\end{equation}
 the best fit result and the dilaton prediction, which are compatible with each other.
 The first uncertainty arises from the lattice data, and the second estimates
  the model-uncertainty
 by varying  the effective $\sig$-mass. Additionally,
different background parametrisations confirm the robustness of this result, and although extra terms can partially mimic the role of the $\sig$-pole, the extracted residue remains stable within uncertainties.

For the pion,  the fits   show firm agreement with the model-independent
soft-pion theorem \eqref{eq:low2}. As in the case of the nucleon the fitted residue
also shows good agreement with the dilaton effective theory.  This time, \EQs\eqref{eq:rPIfit}
and \eqref{eq:rPItheory}  are
\begin{equation}
\label{eq:pi}
\rE{\sig}^\pi = 0.53(16)(3) \;, \qquad
\res{\sig}{\pi}|_{\text{dilaton}} = \frac{2}{3} \pm 0.1 \;,
\end{equation}
the best fit result and the dilaton prediction.  The uncertainties are
of the same type as for the nucleon.
Importantly, while the data are consistent with a dilaton-like residue, the converse, establishing
$\sig$-dominance directly from the lattice results, is not feasible.
This might well be due to the data and the somewhat small residue of the pion as a Goldstone boson,
 and does not undermine the overall consistency of the dilaton interpretation.

Overall, our findings support the idea that QCD is governed by an infrared fixed point
with the $\sig$-meson becoming a light or massless dilaton (in the limit of vanishing light quark masses).\footnote{{If the dilaton Goldberger–Treiman mechanism \eqref{eq:GT} is realised, one can test whether the pion and the sigma decay constants satisfy $F_\pi \approx F_\sig$ 
 in the chiral limit. This relation is essential for incorporating a Yukawa-type mechanism into a strongly coupled sector 
 with a dilaton that mimics the behaviour of the Higgs boson \cite{Cata:2018wzl,Zwicky:2023krx}. 
With $\bar{F}_ \pi  = 70(3) \MeV$  in the $SU(3)$ chiral limit \cite{Bali:2022qja} and assuming $\bar{F}_\pi =  \bar{F}_\sig$, 
one finds  $g_{\sig NN} =  (\bar m_N - \De m_N)/  \bar{F}_\sig \approx 10$. 
This falls well within the phenomenological range reported in the literature \cite{Wu:2023uva} and is therefore deserving of further study,}}
Presumably, the $\sig$-meson  plays a key role in understanding the gravitational $D$-form factor.
In its purest form, with a massless $\sig$, the nucleon form factor develops a pole in the infrared
$D^N(q^2) =  \frac{4}{3}\frac{m_N^2}{q^2} + \ORD(1)$.  For finite $\sig$-mass, relevant to the real world with nonzero quark mass, our analysis in \SEC\ref{sec:note} implies that
$D^N(0) < 0$ holds, provided that the $\sig$-term dominates.

As seen from  \EQs \eqref{eq:N} and \eqref{eq:pi},  the dominant uncertainties stem from the data.
{Further progress can be achieved by improving the precision of the data, extending the kinematic $q^2$-range and
by going to lower quark masses.  Other  potential directions,
are simulating with degenerate quarks (as in beyond-the-Standard-Model studies)
 or dispersive approaches \cite{Cao:2024zlf,Cao:2025dkv}.
  In the longer term we might hope for competitive information
from experiment, including the future electron-ion collider \cite{Accardi:2012qut}.
Another avenue is testing other systems, varying the spin, for which
$\rho$-meson and the $\De$-baryon  gluonic gravitational form factors are available at
$m_\pi \approx 450 \MeV$ \cite{Pefkou:2021fni}. This comes with its very own set of challenges,
partly related to the proton mass decomposition \cite{Ji:1994av}, the $A$-form factor at zero momentum transfer
(extensively studied on the lattice \cite{Borsanyi:2020bpd,Wang:2021vqy,Liu:2021lke,Alexandrou:2024zvn,Alexandrou:2024ozj}),
and will therefore be discussed in a forthcoming papers \cite{Stegeman:2025tdl,appear}.

\paragraph{Acknowledgements:}
RS and RZ are supported by the STFC via the consolidated grants ST/T000600/1 and ST/X000494/1.
We are grateful to Xiong-Hui Cao, 
Martin Hoferichter, Max Hansen,  Xiangdong Ji, Keh-Fei  Liu,  Feng-Kun Guo, Mannque Rho,  David Schaich
and Hanquing Zheng for discussions.
{We thank Wojciech Broniowski and Enrique Ruiz Arriola for sharing the correlations of their fit parameters in \REF\cite{Broniowski:2024oyk}, and the members of the $\chi$QCD collaboration for sharing the data of \REF\cite{Wang:2024lrm}.}
A very special thanks goes to  the MIT-group members Daniel Hackett,
Patrick Oare, Phiala Shanahan and especially Dimitra Pefkou for providing us with their form factor data~\cite{Hackett:2023nkr,Hackett:2023rif}.
Analytic computations were performed with FeynCalc \cite{Shtabovenko:2023idz} and
some Passarino-Veltman functions were numerically checked with LoopTools  \cite{Hahn:1998yk} for real kinematics.

\appendix

\section{The Gell-Mann L\'evy \LsM - a Toy Model}
\label{app:LsigM}

We use the \LsM   \cite{Gell-Mann:1960mvl} as a toy model to illustrate the treatment of particles with a sizeable width, thereby providing a basis for interpreting the fit ansatz of Eq.~\eqref{eq:D}.
Prior to discussing the underlying physics of the \LsM and the details of the corresponding calculation, we first introduce the relevant nucleon form factor $F(q^2)$\footnote{We may regard  this form factor as a toy model for the improvement term $T_{\mu \nu } \supset  \frac{1}{3} (\partial_\mu \partial_\nu - \eta_{\mu\nu} \partial^2)  \sig $ \eqref{eq:TRdef}.}
  \begin{equation}
  \label{eq:F}
  \matel{N(p')}{\sig}{N(p)} = \frac{1}{2m_N^2} \bar u(p') F(q^2) u(p) \;, \quad q= p'-p \;.
  \end{equation}
 The LO expression  is governed  by  $g = g_{\sig NN}$, the coupling of the $\sig$ to two nucleons,
  \begin{equation}
 F(s) |_{\text{LO}}= \frac{2 m_N^2 g }{s - m_\sig^2} \;.
  \end{equation}
     The NLO form factor can be parametrised as
  \begin{equation}
  \label{eq:FNLO}
  F(s) |_{\text{NLO}}= \frac{2 m_N^2 g ( 1+ \Fone(s)) } {s - m_\sig^2 -  \Sig(s)} \;,
  \end{equation}
  where $\Sig(s)$ and $\Fone(s)$ are the self-energy  and the vertex corrections, respectively.

In the following, we define the model, present the NLO computation, and carry out the analytic continuation.  We then use the result to assess the impact of the broad $\sigma$-resonance on the form factor parametrisation.

\subsection{The \LsM Lagrangian in the broken phase}
\label{app:Lag}

 The \LsM is a most formidable model, first introduced by Schwinger \cite{Schwinger:1957em} to implement $SU(2)_L \times SU(2)_R$ invariance in the strong-interaction sector.
  It was then refined by
 Gell-Mann and L\'evy  \cite{Gell-Mann:1960mvl} to include the mechanism of
 spontaneous symmetry breaking, which
 leads to massless pions and mass generation of the nucleon,  referred to  as the Yukawa-mechanism.
The renormalisation was worked out in \REF\cite{Gervais:1969zz}, providing an important precursor to the renormalisation of the electroweak sector of the Standard Model.
It should be emphasised that the $\sig$-particle, in this context, is not to be confused with the $\sig$-meson of dilaton chiral perturbation theory. In fact, integrating out a potentially heavy $\sig$, in the linear model, leads to chiral perturbation theory, i.e. the \emph{non-linear} $\sig$-model.
This allows to determine the low energy constants explicitly.
 Gasser and Leutwyler  \cite{Gasser:1983yg} make the point that, since the predictions differ from the values observed in nature, the \LsM\ cannot be considered a viable theory of the strong interactions.

The Gell-Mann L\'evy linear  $\sig$-model with pion mass  \cite{Gell-Mann:1960mvl}, reads
 \begin{alignat}{2}
& \Lag  &\;=\;&  \frac{1}{2} (\partial \pi)^2 +  \frac{1}{2} (\partial \sig)^2   - \frac{1}{2} m_\sig^2 \sig^2
 + \bar N ( i \slashed{\partial} - g (\sig - i \pi \ga_5)) N   - V(\sig, \pi) \;,
\end{alignat}
 where $\pi = \pi^a T^a$ is understood in the pion to nucleon coupling and the
 potential is given by
 $V(\sig, \pi)  =  \la/4 (\sig^2 + \pi^2 - v^2)^2- H \sig$,  where $\pi^2 = \pi^a \pi^a$,
 and $H$ is the (pion) mass perturbation. The potential is minimised by
 $\partial_\sig V(F_\pi, 0) =0$, where $F_\pi$ differs from $v$ by the perturbation
  \begin{equation}
 F_\pi^2 = v^2 + \frac{H}{\la F_\pi} \;.
 \end{equation}
 The vacuum expectation value $\vev{\sig} = F_\pi$ spontaneously breaks the global symmetry $SU(2)_L \times SU(2)_R$ down to the isospin subgroup $SU(2)_V$, with the pions emerging as the associated Goldstone bosons. The constant $F_\pi$ is identified with the pion decay constant, defined through the axial current matrix element $\matel{0}{A^a_\mu}{\pi^b} = i p_\mu F_\pi$.
 The  new  potential then reads
 \begin{equation}
  V(\sig+F_\pi,\pi)   =  \frac{\la}{4} ( (\sig + F_\pi)^2 + \pi^2 - v^2)^2 -  H(\sig + F_\pi) \;,
 \end{equation}
from where the masses
  \begin{alignat}{3}
  \label{eq:mpi}
&   m_\pi^2  &\;=\;&  \la(F_\pi^2 - v^2) &\;=\;& \frac{H}{F_\pi}    \;,\nonumber \\[0.1cm]
 &  m_\sig^2 &\;=\;& \la(3 F_\pi^2 - v^2)
 &\;=\;&  2 \la F_\pi^2 + m_\pi^2   \;,\nonumber \\[0.1cm]
 & m_N^2 &\;=\;&  g^2  F_\pi^2   \;, & &
 \end{alignat}
 and couplings ($V \supset \la_n \sig^n$)
 \begin{equation}
 \label{eq:coup}
\la_4 = \frac{\la}{4} \;,\quad  \la_3 = \la F_\pi  \;, \quad  g_{\sig NN} =  - g_{\pi NN} = g   \;,
 \end{equation}
can be read off.
We see that the nucleon mass is generated by a Yukawa mechanism, the pion mass by the explicit
breaking term $H$, and the $\sig$ mass is governed by $\la$ and the explicit breaking.
The model has four free parameters which we choose to be
\begin{equation}
\label{eq:free}
( g,\la,m_\pi,F_\pi)  \;.
\end{equation}
The renormalisation  \cite{Gervais:1969zz} of the \LsM is involved, but since we do not aim
to match it to experiment we may  adapt the $\MSbar$-scheme within dimensional
regularisation.  In fact, we slightly modify it for the self-energies to cancel tadpole diagrams
 (for $\mu = m_\sig$), so that we may simply ignore them.  The concrete values used for  \eqref{eq:free} will be discussed in  \SEC\ref{app:broad}.

\subsection{Relevant NLO corrections}
\label{app:NLO}

Our next task is to compute the NLO corrections \eqref{eq:FNLO}.
This includes the self-energy and the vertex correction
which determine the pole position and the complex residue, respectively.
The results are evaluated in terms of Passarino-Veltman functions, defined by
\begin{equation*}
\label{eq:PaVeConvention}
I_n(\ell_1^2, \ell_2^2, \dots) = \frac{\mu^{4-d}}{i \pi^{\frac{d}{2}}} \int  \frac{d ^dk }{(k^2 - m_0^2 +
  i 0)((k+\ell_1)^2 - m_1^2  +i 0)((k+\ell_1 + \ell_2)^2 - m_2^2 +i 0) \dots } \;,
\end{equation*}
where  $A_0 = I_0$, $B_0 = I_1 $ and $  C_0= I_2$. For the  momentum routing
of the triangle function $C_0$ we use the   LoopTools   conventions  \cite{Hahn:1998yk}.

 \subsubsection{Self-energy corrections $\Sig(s)$}
 \label{app:self}

The goal of this section is to determine the $\sig$-pole, on the second sheet,
 from the (inverse) propagator
\begin{equation}
\label{eq:inverse}
\De^{-1}(s) = {s- m_\sig^2 - \Sig(s)} \;.
\end{equation}
We decompose  the self-energy into parts
\begin{equation}
\Sig(s) = \Sig_{\la_3}(s) + \Sig_{\la_4}(s)+ \Sig_{g}(s) \;,
\end{equation}
proportional to the couplings
$\la_{3}$, $\la_4$ and $g$ as given in \EQ\eqref{eq:coup}, with the $\Sig_{\la_3,g}$-contributions shown
 in \FIG\ref{fig:linsig}.
An explicit evaluation  of all NLO diagrams gives
 \begin{alignat}{2}
 \label{eq:SigNLO}
&  \Sig_{\la_3}(s) &\;=\;&  \frac{( \la_3)^2 }{16 \pi^2}   (  -  2 (N_f^2-1)    B_0(s,m_\pi^2,m_\pi^2) - 18 B_0(s,m_\sig^2,m_\sig^2) )  \;,
   \nonumber \;  \\
   &  \Sig_{\la_4} (s)    &\;=\;&  \frac{   \la_4 }{16 \pi^{2}} (-12 A_0(m_\sig^2)   -  2(N_f^2-1)A_0(m_\pi^2))
       \;, \nonumber  \\
 &  \Sig_{g}(s)     &\;=\;&    \frac{ g^2 }{16 \pi^{2}} 4 N_f \left( ( 2 m_N^2-s/2) B_0(s,m_N^2,m_N^2)  +  A_0(m_N^2) \right)  \; ,
 \end{alignat}
where the number of fermions is $N_f =2$ (proton and  neutron).

 \begin{figure}[h]
\centering
\begin{tikzpicture}[scale=1.2, transform shape]

  \begin{scope}[shift={(0,0)}]
    \begin{feynman}
      \diagram [horizontal=a to c] {
        i1 [particle=$\sigma$] -- [scalar, line width=1.5pt] a -- [scalar, line width=1.5pt, draw=white] c -- [scalar, line width=1.5pt] f1 [particle=$\sigma$],
      };
      \draw[scalar, line width=1.5pt] (a) to[out=90, in=90, looseness=1.5] (c);
      \draw[scalar, line width=1.5pt] (c) to[out=-90, in=-90, looseness=1.5] node[midway, below, color=white] {$\pi$} node[midway, below=15pt] {\scriptsize (A)} (a);
    \end{feynman}
  \end{scope}

  \begin{scope}[shift={(0.25\textwidth,0)}]
    \begin{feynman}
      \diagram [horizontal=a to c] {
        i1 [particle=$\sigma$] -- [scalar, line width=1.5pt] a -- [scalar, line width=1.5pt, draw=white] c -- [scalar, line width=1.5pt] f1 [particle=$\sigma$],
      };
      \draw[thick] (a) to[out=90, in=90, looseness=1.5] (c);
      \draw[thick] (c) to[out=-90, in=-90, looseness=1.5] node[midway, below] {$\pi$} node[midway, below=15pt] {\scriptsize (B)} (a);
    \end{feynman}
  \end{scope}

  \begin{scope}[shift={(0.5\textwidth,0)}]
    \begin{feynman}
      \diagram [horizontal=a to c] {
        i1 [particle=$\sigma$] -- [scalar, line width=1.5pt] a -- [scalar, line width=1.5pt, draw=white] c -- [scalar, line width=1.5pt] f1 [particle=$\sigma$],
      };
      \draw[double, double distance=1.5pt, thick] (a) to[out=90, in=90, looseness=1.5] (c);
      \draw[double, double distance=1.5pt, thick] (c) to[out=-90, in=-90, looseness=1.5] node[midway, below] {$N$} node[midway, below=15pt] {\scriptsize (C)} (a);
    \end{feynman}
  \end{scope}

\end{tikzpicture}
\caption{\small Diagrams for the self-energy corrections $\Sig_{\la_3,g}$.
The two diagrams for $\Sig_{\la_4}$ are not shown.
}
\label{fig:linsig}
\end{figure}
We start by expressing \eqref{eq:inverse} directly in terms of renormalised quantities,
\begin{equation}
\label{eq:Deinv}
 \De^{-1}(s) = s -   m_\sig^2 - \bar{\Sig}_{\la_3}(s) - \bar{\Sig}_{\la_4}(s) - \bar{\Sig}_{g}(s)  \;,
\end{equation}
where the bars indicate that we work in the $\MSbar$ scheme without tadpoles, meaning that $A_0,B_0$ in \eqref{eq:SigNLO} are replaced by
\begin{equation}
  \bar{A}_0(m_a^2) = m_a^2 \bigg( 1 - \ln \frac{m_a^2}{\mu^2} \bigg) \;, \qquad
  \bar{B}_0(s,m_a^2,m_a^2) = 2 - \ln \frac{m_a^2}{\mu^2} - \be_a \ln \!\bigg( \frac{\be_a+1}{\be_a-1} \bigg) \;,
\end{equation}
with $\be_a = \sqrt{1 - 4m_a^2/s}$, as usual.
The $\sig$-pole is then obtained by analytically continuing the inverse propagator to the second sheet,
\begin{equation}
\label{eq:ana2}
 \big( \De^{(II)}(s_\sig) \big)^{-1} = 0 \;,
\end{equation}
where continuity across the cut is imposed via
$ (\De^{-1})^{(II)}(s-i0) = \De^{-1}(s+i0) $.
Among the loop functions, only the pion $\bar B_0$ requires a non-trivial continuation
\begin{equation}
\bar{B}^{(II)}_0(s,m_\pi^2,m_\pi^2) =
2 - \ln \frac{m_\pi^2}{\mu^2} - \be_\pi(s) \ln \frac{1+\be_\pi(s)}{1-\be_\pi(s)} - i\pi \;,
\end{equation}
which contributes a large part  the imaginary part of $s_\sig$.
Equation \eqref{eq:ana2} can be solved numerically, or alternatively approximated perturbatively
by substituting $s = m_\sig^2$ into the self-energy\footnote{This expression coincides with Eq.~(1) in \cite{Masjuan:2008cp}, up to the choice of scheme and with the $\Sig_g$ term omitted.}
\begin{equation}
\label{eq:ssig2}
s_\sig = m_\sig^2 + \bar{\Sig}^{(II)}_{\la_3}(m_\sig^2) + \bar{\Sig}^{(II)}_{\la_4}(m_\sig^2) +
 \bar{\Sig}^{(II)}_{g}(m_\sig^2) \;.
\end{equation}
In practice, we use the numerical solution of \eqref{eq:Deinv} on the second sheet, which generally yields results close to the analytic approximation.

 \subsubsection{Vertex corrections $\Fone(s)$}
 \label{app:vertex}

The   diagrams  for the vertex corrections, in the  order  shown  in  \FIG\ref{fig:vertex}, are
\begin{alignat}{2}
&  \matel{N}{\sig}{N}|_{NLO}     &\;=\;&  \frac{ 6i g^2  \la_3}{s-m_\sig^2} \int \frac{d^4 k}{(2 \pi)^4}
{\bar u(p')  S_N(k)  u(p) } \De_\sig(k-p')\De_\sig(k-p) \;
   \nonumber \;  \\
 &      &\;+\;&   \frac{ -6i  g^2 \la_3}{s-m_\sig^2} \int \frac{d^4 k}{(2 \pi)^4}
{\bar u(p')   \ga_5 S_N(k)   \ga_5  u(p) } \De_\pi(k-p')\De_\pi(k-p)  \;  \nonumber \;  \\
&      &\;+\;&   \frac{ i g^3  }{s-m_\sig^2} \int \frac{d^4 k}{(2 \pi)^4}
{\bar u(p')  S_N(k-p) S_N(k-p')  u(p) }  \De_\sig(k) \;  \;.
   \,
\end{alignat}
with standard scalar and fermion propagators   $\De_\sig(k) = \frac{1}{k^2 - m_\sig^2}$ and $S_N(k) = \frac{\slashed{k} + m_N}{k^2-m_N^2}$.
The form factor or vertex corrections are obtained
 by matching to \eqref{eq:F}
\begin{equation}
F(s) =   {2 m_N^2} \matel{N}{\sig}{N}|_{\bar uu}  \;,  \quad
\Fone(s) =   \frac{1}{g} ( s-m_\sig^2)   \matel{N}{\sig}{N} |_{\bar uu}  \;,
\end{equation}
where $|_{\bar uu}$ denotes the projection onto the spinors.
We find
\begin{alignat}{2}
\label{eq:pragma}
 &16 \pi^2  \Fone(s)    &\;=\;&   \frac{6 \la m_N^2}{s- 4 m_N^2}
 \left( P_-(m_\sig^2)  C_0(s,m_\sig^2)  + 2
\Delta  B_0(s, m_\sig^2,m_\sig^2)  \right) \;    \nonumber \\[0.1cm]
 & &\;+\;&   \frac{6 \la m_N^2 }{s- 4 m_N^2}
  \left( P_+(m_\pi^2) C_0(s,m_\pi^2)  + 2 \De B_0(s,m_\pi^2,m_\pi^2) \right)    \;     \nonumber \\[0.1cm]
 & &\;-\;&
  g^2   (B_0(s,m_N^2,m_N^2)  +
m_\sigma^2 \, c_0(s,m_\sig^2))   \;  \;,
\end{alignat}
with $P_{\pm}(\Mass^2) =  ( 4 m_N^2- 2 \Mass^2)  \pm   (s-4 m_N^2)$ making the chirality structure manifest,
\begin{equation}
 \De B_0(s, m_\sig^2,m_N^2) = B_0(s, m_\sig^2,m_\sig^2) -
  B_0 (m_N^2,m_N^2,m_\sig^2) \;,
\end{equation}
which is an ultraviolet-finite contribution,
and the abbreviated triangle functions are given by
\begin{equation*}
C_0(s,\Mass^2) = C_0(s,m_N^2, m_N^2,\Mass^2,\Mass^2,m_N^2) \;, \quad
c_0(s,\Mass^2) =  C_0(s,m_N^2, m_N^2, m_N^2, m_N^2, \Mass^2) \;.
\end{equation*}
Note that one of the $g$-factors gets absorbed into $m_N = g v$ in the first two contributions
and  the singularity  at $s=4m_N^2$ is only apparent; an artefact of the Passarino-Veltman reduction.
This means that
\begin{equation}
(4 m_N^2- 2 m_\sig^2) C_0(s,m_\sig^2) + 2(
 B_0(s, m_\sig^2,m_\sig^2) -
  B_0 (m_N^2,m_N^2,m_\sig^2))\Big|_{s = 4 m_N^2} = 0 \;,
\end{equation}
 must hold which we  checked analytically for the imaginary part (using the expressions in the next section)
  and numerically for the real part.
\begin{figure}[h]
\centering
\begin{minipage}{0.3\textwidth}
\centering
\begin{tikzpicture}[scale=1.2, transform shape]
\begin{feynman}
\vertex (c1) at (0,0);
\vertex (b3) at (0.75,0);
\vertex (b1) at (1.5,0.5);
\vertex (b2) at (1.5,-0.5);
\vertex (a1) at (2.2,0.5) {$N$};
\vertex (a2) at (2.2,-0.5) {$N$};
\node at (-0.3,0) {$\sigma$};
\node at (1.125,-1.2) {\scriptsize (A)};
\filldraw[black] ($(c1)+(-3pt,-3pt)$) rectangle (3pt,3pt);
\diagram* {
  (c1) -- [scalar, line width=1.5pt] (b3),
  (b3) -- [scalar, line width=1.5pt] (b1),
  (b3) -- [scalar, line width=1.5pt] (b2),
  (b1) -- [thick,double, double distance=1.5pt] (a1),
  (b2) -- [thick,double, double distance=1.5pt] (a2),
  (b1) -- [thick,double, double distance=1.5pt] (b2),
};
\end{feynman}
\end{tikzpicture}
\end{minipage}
\begin{minipage}{0.3\textwidth}
\centering
\begin{tikzpicture}[scale=1.2, transform shape]
\begin{feynman}
\vertex (c1) at (0,0);
\vertex (b3) at (0.75,0);
\vertex (b1) at (1.5,0.5);
\vertex (b2) at (1.5,-0.5);
\vertex (a1) at (2.2,0.5) {$N$};
\vertex (a2) at (2.2,-0.5) {$N$};
\node at (-0.3,0) {$\sigma$};
\filldraw[black] ($(c1)+(-3pt,-3pt)$) rectangle (3pt,3pt);
\node at (1.125,-0.5) {$\pi$};
\node at (1.125,-1.2) {\scriptsize (B)};
\diagram* {
  (c1) -- [scalar, line width=1.5pt] (b3),
  (b3) -- [thick] (b1),
  (b3) -- [thick] (b2),
  (b1) -- [thick,double, double distance=1.5pt] (a1),
  (b2) -- [thick,double, double distance=1.5pt] (a2),
  (b1) -- [thick,double, double distance=1.5pt] (b2),
};
\end{feynman}
\end{tikzpicture}
\end{minipage}
\begin{minipage}{0.3\textwidth}
\centering
\begin{tikzpicture}[scale=1.2, transform shape]
\begin{feynman}
\vertex (c1) at (0,0);
\vertex (b3) at (0.75,0);
\vertex (b1) at (1.5,0.5);
\vertex (b2) at (1.5,-0.5);
\vertex (a1) at (2.2,0.5) {$N$};
\vertex (a2) at (2.2,-0.5) {$N$};
\node at (-0.3,0) {$\sigma$};
\node at (1.125,-1.2) {\scriptsize (C)};
\filldraw[black] ($(c1)+(-3pt,-3pt)$) rectangle (3pt,3pt);
\diagram* {
  (c1) -- [scalar, line width=1.5pt] (b3),
  (b3) -- [thick,double, double distance=1.5pt] (b1),
  (b3) -- [thick,double, double distance=1.5pt] (b2),
  (b1) -- [thick,double, double distance=1.5pt] (a1),
  (b2) -- [thick,double, double distance=1.5pt] (a2),
  (b1) -- [scalar, line width=1.5pt] (b2),
};
\end{feynman}
\end{tikzpicture}
\end{minipage}
\caption{\small Feynman diagrams for the NLO corrections to the $\sigma NN$ vertex.}
\label{fig:vertex}
\end{figure}

\subsection{Second-sheet analytic continuation of the form factor}
\label{app:2nd}

In order to deduce the complex residue we must analytically continue
the expression in \eqref{eq:pragma} to the second sheet where the pole \eqref{eq:ssig2} lies.
The generic analytic continuation for a  function $f(s)$, defined on the first sheet, reads
    \begin{equation}
    \label{eq:2nd}
f^{(II)}(s)  = f(s) + \text{disc}[ f(s)]  \;,
  \end{equation}
  where $\text{disc}[ f(s)]  = f(s+i0) - f(s-i0)$  ensures continuity across the cut, $f^{(II)}(s-i0) = f(s+i0)$.
{Concretely, we have the imaginary parts of the Passarino-Veltman functions analytically,
   while the real parts can be evaluated numerically through the LoopTools package  \cite{Hahn:1998yk} (for the $C_0$ function).} Fortunately, this proves sufficient since
  i)   the form factors satisfy real analyticity which implies
  $\text{disc}[ f(s)]  = 2 i \Ima f(s)$ by  Schwartz's reflection principle and ii) a representation for
  $f(s)$ valid on the physical sheet is given by the standard dispersion relation
\begin{equation}
\label{eq:master}
f^{(II)}(s) = \frac{1}{\pi} \int ds' \frac{\Ima f(s')}{s'-s-i0} + 2 i  \Ima f(s)  \;.
\end{equation}
This  expression serves as the master formula for analytic continuation.

The imaginary parts of $B_0$ and $C_0$
 are provided in Itzykson and Zuber \cite{Itzykson:1980rh}, although the expression for $C_0$  contains significant typos in the Källén function, which we corrected and verified against LoopTools.
On the real line we express them as
 \begin{alignat}{2}
 \label{eq:Im}
&  \Ima B_0(s, m_1^2,m_2^2)
 &\;=\;&  \pi \frac{\sqrt{\la_m }}{s} \theta(s- (m_1+m_2)^2) \;, \\[0.1cm]
& \Ima C_0(s,p_1^2,p_2^2, m_1^2,m_2^2,m_3^2)   &\;=\;&     \frac{  - \pi}{\sqrt{\la_p} }  \ln \frac{a+b}{a-b}\theta(s- (m_1+m_2)^2)  \;,
     \end{alignat}
 where  $\la_p = \la(s,p_1^2,p_2^2)$, $\la_m = \la(s,m_1^2,m_2^2)$
 and
 $ \la  (s,m_1^2,m_2^2) = (s - (m_1-m_2)^2) (s - (m_1+m_2)^2) $ is the K\"all\'en function,  and
    \begin{equation}
a = s^2  - s( p_1^2+p_2^2 +m_1^2+m_2^2 - 2 m_3^2) - (p_1^2-p_2^2)(m_1^2 - m_2^2) \;,
\quad
  b = \sqrt{\la_p} \sqrt{\la_m}  \;.
   \end{equation}
   For equal masses one has $\Ima B_0(s,m_a^2,m_a^2) = \pi  \be_a \theta(s-4 m_a^2)$ which is most often required.

Concretely, for $C_0(s,\Mass^2)$ and similarly for $c_0(s,\Mass^2)$,
\begin{equation}
C_0^{(II)}(s,\Mass^2) =  \int_{4 \Mass^2}^\infty \frac{ds' \frac{1}{\pi} \Ima C_0(s',\Mass^2) }{s'-s}  +
2i  \Ima C_0(s,\Mass^2) \;,
\end{equation}
gives a formula valid  on the second sheet (away from the cut).
For the second type of term
we need a once-subtracted dispersion relation
\begin{equation}
\De B^{(II)}_0(s,\Mass^2,\Mass^2) = \De B_0(s_0,\Mass^2,\Mass^2)
+ (s-s_0)  \int_{4 \Mass^2}^\infty \frac{ds' \frac{1}{\pi} \Ima B_0(s',\Mass^2,\Mass^2) }{(s'-s)(s'-s_0)}  +
2i  \Ima B_0(s,\Mass^2,\Mass^2) \;.
\end{equation}
where $\Ima \De B_0(s,\Mass^2,\Mass^2) =   \Ima B_0(s,\Mass^2,\Mass^2)$ has been used.
We will
choose the subtraction point   $s_0 = 0$ below the cut
$ 4 \Mass^2$ ($\Mass^2 = m_{\pi,\sig}^2$).  A similar formula is applied for $B_0(s,m_N^2,m_N^2) $
which will be shown explicitly in the final result.
 It should be stressed that the imaginary parts only need to be added for the pion loops since the sigma loops are above the point where $s$ is continued to the second sheet.
The explicit subtraction constants are  given by
    \begin{equation}
  \label{eq:above}
  \De B_0 ( 0,m^2,m_N^2) =    \int_0^1 dx \ln  (\frac{ m_N^2}{m^2} x^2+ (1-x) ) \;, \quad
  \bar{B}_0(0, m^2, m^2) = -\ln \frac{m^2}{\mu^2} \;.
\end{equation}
 We note that the $\mu$-dependence in $\De B_0$ vanishes between the two terms
since the difference is ultraviolet-finite. For $\bar{B}_0$, the $\mu$-dependence remains and can be used to assess the uncertainty.
The analytic continuation of  \eqref{eq:pragma}, valid on the entire second sheet,
reads
 \begin{alignat}{2}
\label{eq:pragma2}
 &16 \pi^2  \Fone^{(II)}(s)     =    \frac{6 \la m_N^2}{s- 4 m_N^2}
 \Big( \!\!\! &2&  \big( \De B_0(0,m_\sig^2,m_N^2)  +
    s \int_{4 m_\sig^2}^\infty \frac{ds' \frac{1}{\pi} \Ima B_0(s',m_\sig^2,m_\sig^2) }{s'(s'-s)} \big)    \\[0.1cm]
 & &\;+\;&   P_-(m_\sig^2) \;  \int_{4 m_\sig^2}^\infty \frac{ds' \frac{1}{\pi} \Ima C_0(s',m_\sig^2) }{s'-s}
    +   \{ m_\sig \to m_\pi, P_- \to P_+ \}     \nonumber   \\[0.1cm]
&   &\;+\;&  i \big[ 4 \Ima B_0(s,m_\pi^2,m_\pi^2)    + 2 P_+(m_\pi^2) \Ima  C_0(s,m_\pi^2)
   \big] \Big)  \nonumber   \\[0.1cm]
 &     \quad\; - g^2   \Big(    \bar B_0(0,m_N^2 ,m_N^2) \;+ &\;s\;&  \! \int_{4 m_N^2}^\infty \frac{ds' \frac{1}{\pi} \Ima B_0(s',m_N^2,m_N^2) }{s'(s'-s)} + m_\sig^2 \int_{4 m_N^2}^\infty \frac{ds' \frac{1}{\pi} \Ima c _0(s',m_\sig^2) }{s'-s} \Big)
      \; , \nonumber
 \end{alignat}
with the constant functions as per \eqref{eq:above} and the previously given imaginary parts above.
The term in square brackets is due to analytic continuation across the two-pion threshold.

\subsection{Concluding the form-factor study in the \LsM}
\label{app:broad}

 We now return to the investigation of the form factor in the \LsM  and assess whether the Euclidean ansatz in \eqref{eq:D} provides a reliable description, and to what extent the complex residue $r_\sig$ differs from its Euclidean counterpart $\rE{\sig}$.
 The  form factor  $F(q^2)$ \eqref{eq:F} satisfies the standard dispersion relation
\begin{equation}
\label{eq:DeLsM}
F(q^2) = \int_{\text{cut}} ds \, \frac{\rho_{F}(s)}{s-q^2-i0} \;,
\end{equation}
with $\rho_{F}(s) = \tfrac{1}{\pi} \Ima F(s)$ due to real analyticity.
The plot in \FIG\ref{fig:LsMDe} demonstrates that the form factor is well approximated by an effective mass $\mE{\sig}$, fitted in the range $q^2 \in [-2.95,0.05] \GeV^2$
\begin{equation}
\label{eq:Feffective}
F(q^2) \big|_{q^2<0}  = 2 m_N^2 g \frac{\rE{\sig}}{q^2-\mE{\sig}^2} \;,
\end{equation}
with $\mE{\sig}$ is close to  the centre of the
$\rho_F$-distribution, as one would expect.
The difference to  the generic case \eqref{eq:Euclid} is that  here there is only the pole contribution.
 In a calculable model, however, such contributions can be separated and are thus not of primary concern.
\begin{figure}
  \includegraphics[width=0.49\textwidth]{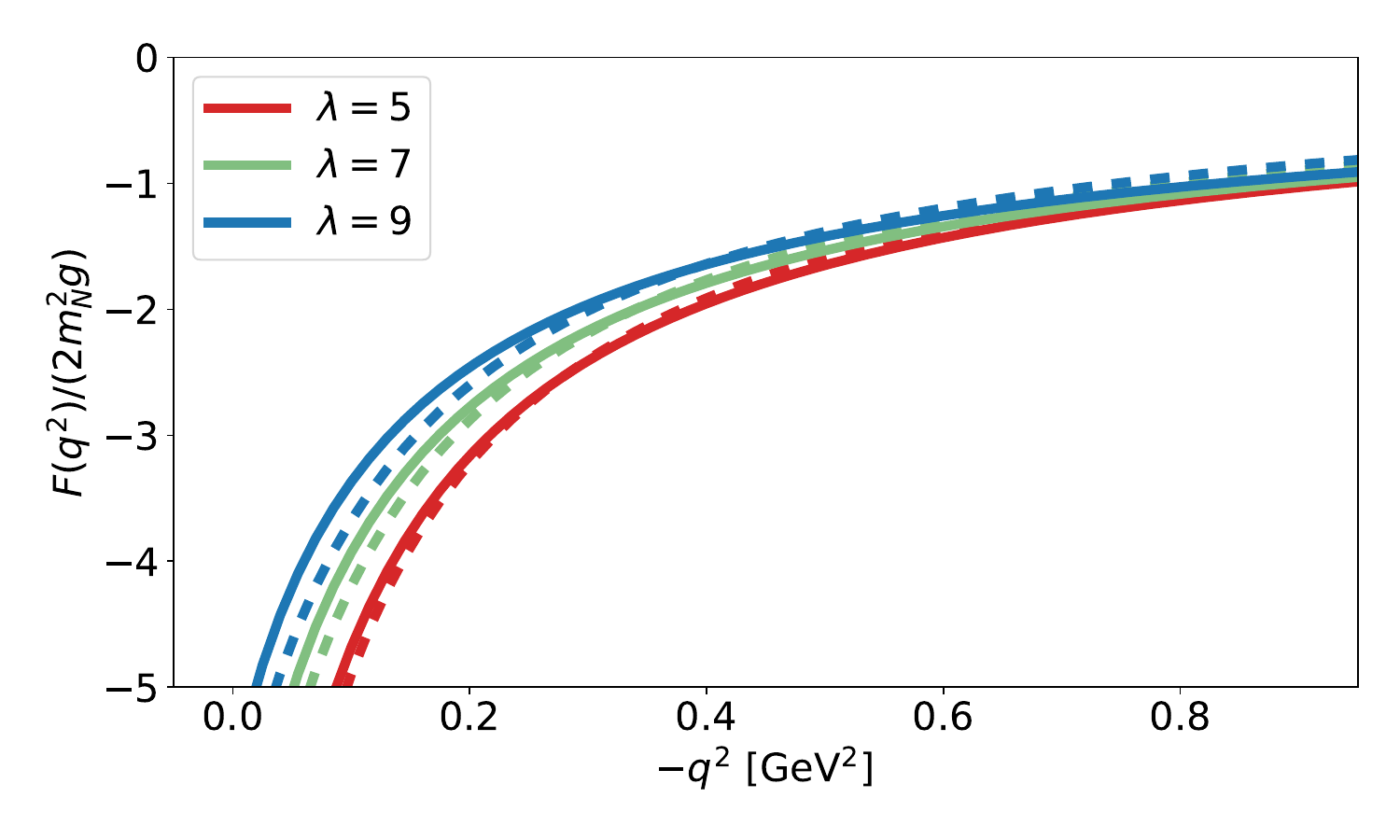}
  \includegraphics[width=0.49\textwidth]{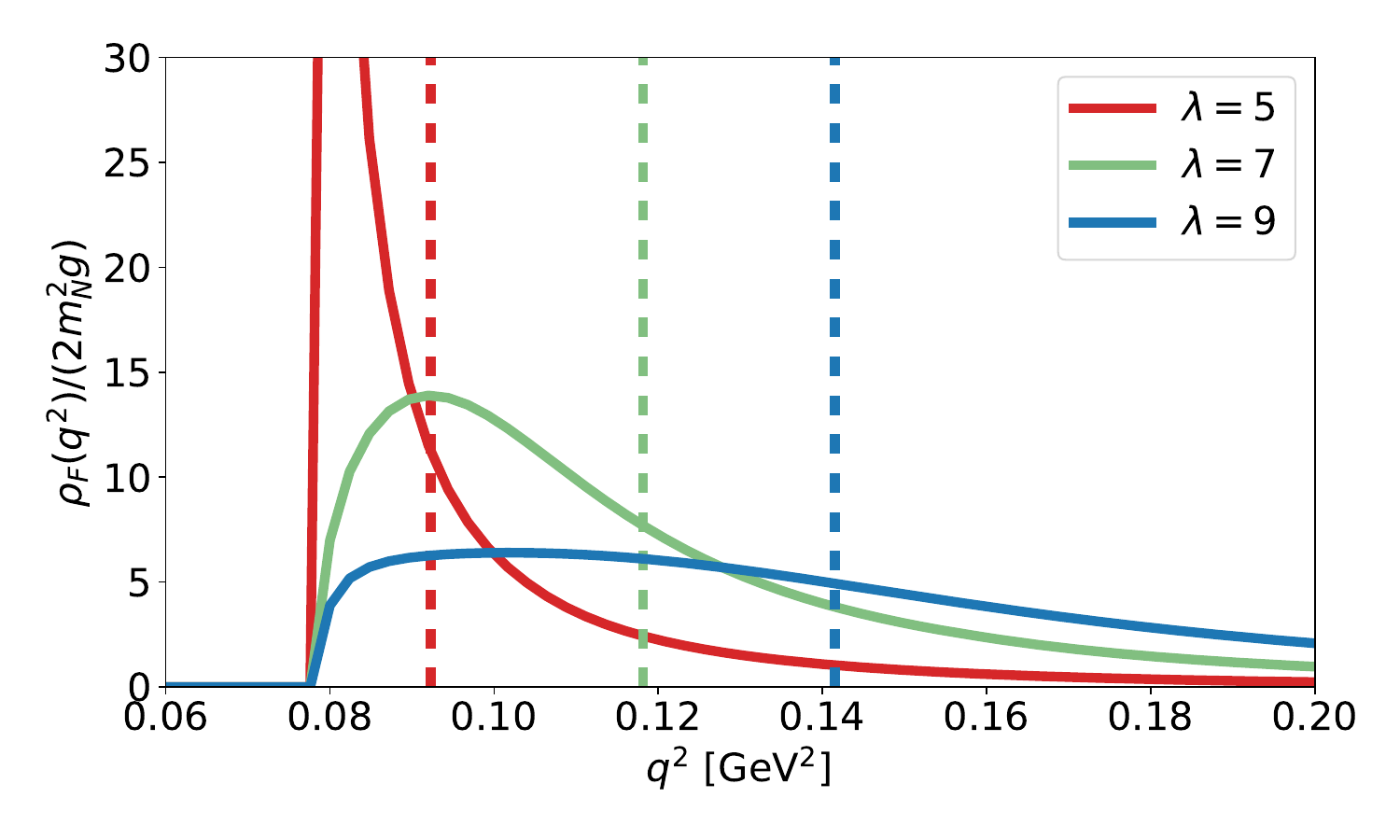}
  \caption{\small The NLO form factor $F(q^2)$ \eqref{eq:FNLO} (left) and
  the corresponding density $\rho_F(s)=\tfrac{1}{\pi}\Ima F(s)$ (right).
  The input values \eqref{eq:free} are $m_\pi=140\MeV$, $F_\pi=93\MeV$ and $\la=(g+1/2)^2$ for
  three different value of $\la$.
   The solid lines corresponds to the analytic  \LsM computation and the dashed lines are
  fitted effective pole representations \eqref{eq:Feffective}.
   The \LsM and effective pole curves are nearly identical in the Euclidean, despite rather different densities in the Minkowski region.}
    \label{fig:LsMDe}
\end{figure}
\begin{table}[h]
\centering
\begin{tabular}{ l c  ||   c c ||   c r }
$\lambda$ &  $m_\sig^{\text{LO}}{\small [\GeV]}$     & $m_{E,\sigma}{\small [\GeV]}$ & $\de r_{E,\sigma}^{\text{LO}}$
& $\sqrt{s}_\sigma {\small [\GeV]}$     & $\de r_\sigma $ \\ \hline
3 & 0.27 & 0.25 & $-0.036$ & $ 0.24 e^{0.6^{\circ} i}$ & $ -0.13 + 0.03 i $ \\
5 & 0.33 & 0.30 &$ -0.059$ & $ 0.28 e^{-4.8^{\circ} i}$ & $ -0.07 + 0.02 i $ \\
7 & 0.38 & 0.34 & $-0.084$ & $ 0.30 e^{-9.7^{\circ} i}$ & $ -0.14 - 0.10 i $ \\
9 & 0.42 & 0.38 & $-0.112$ & $ 0.36 e^{-16.1^{\circ} i}$ & $ -0.08 - 0.25 i $ \\
11 & 0.46 & 0.40 & $-0.142$ & $ 0.42 e^{-19.1^{\circ} i}$ & $ -0.00 - 0.34 i $ \\
13 & 0.49 & 0.42 & $-0.174$ & $ 0.49 e^{-20.7^{\circ} i}$ & $ 0.07 - 0.40 i $ \\
15 & 0.53 & 0.44 &$ -0.208$ & $ 0.55 e^{-21.6^{\circ} i}$ & $ 0.14 - 0.45 i $ \\
\end{tabular}
\caption{\small
Comparison of the Euclidean and complex pole parameters as a function of $\la = (g+1/2)^2$.
The complex pole is given in polar coordinates, facilitating comparison with the tree-level mass and the Euclidean mass $m_{E,\sigma}$.
The residues can be reconstructed by adding unity according to \eqref{eq:der}.
The Euclidean values are obtained from fits to form-factor data computed in the \LsM\ at 20 evenly spaced points in the interval $q^2 \in [-2.95,0.05]\GeV^2$.
}
\label{tab:resi}
\end{table}
We next compare the complex residue with the effective Euclidean one.
The former  can be obtained from
\begin{equation}
r_\sig = \lim_{q^2 \to s_\sig} (q^2-s_\sig) F^{(\text{II})}(q^2) = 1 + \Fone^{(II)}(s_\sig) \,,
\end{equation}
where $s_\sig$ is the complex pole and $\Fone^{(II)}$ refers to the analytic continuation \eqref{eq:pragma}.

For illustration we adopt the QCD-inspired values $F_\pi=93\MeV$ and $m_\pi=140\MeV$, while varying $\la=(g+1/2)^2$.
This choice maintains the hierarchy between scalar and fermionic loop contributions.
One can choose other values in the $(\la,g)$-parameter space but as long as there are no strong
cancellations the main characteristics remain unchanged.
In fact, tuning as $\la=(g+1/2)^2$ is in line with softening the high-energy behaviour akin
to the   Regge trajectory  in QCD.

The results, summarised in \TAB\ref{tab:resi}, are expressed in terms of deviations from the normalised LO residue:
\begin{equation}
\label{eq:der}
r_\sig = 2 m_N^2 g \, (1+\de r_\sig) \;,
\qquad
\rE{\sig} = 2 m_N^2 g \, (1+\de \rE{\sig}) \;.
\end{equation}
From \TAB\ref{tab:resi} we observe that the phases of $\sqrt{s_\sig}$ and $\de r_\sig$ are correlated.
At the same time, the ratio $|\de \rE{\sig}/\de r_\sig|$ differs significantly from unity, confirming the expectation expressed in \SEC\ref{sec:G}.
In general, one further finds that $|\de \rE{\sig}/\de r_\sig| < 1$, while the correlation between $|s_\sig|$ and $\mE{\sig}$ holds qualitatively.
We therefore conclude that the complex residue at the pole and the effective Euclidean residue exhibit substantial qualitative differences and should not be compared quantitatively.

\section{A Multipole Expansion in Momentum Space}
\label{app:multipole}

In this appendix, we present a more systematic perspective on the representation given in \EQ\eqref{eq:Euclid}, formulated in the spirit of a multipole expansion. We start by decomposing
the density into
\begin{equation}
\rho_G(s) = \rho_\party(s) + \rho_b(s) \;,
\end{equation}
where $\rho_{\party,b}$ denote the $\party$-resonance and background contributions, respectively.
Next, consider expanding the denominator of the dispersion integral \eqref{eq:Gdis} around $s = \mE{\sig}^2$
\begin{equation}
\frac{1}{s-q^2} = \frac{1}{\mE{\party}^2-q^2} \Bigg( 1 +
\sum_{n \geq 0} \eps^n \Bigg) \;, \qquad
\eps = \frac{\mE{\party}^2 - s}{\mE{\party}^2 - q^2} \;.
\end{equation}
If $\mE{\party}^2$ is chosen at the centre of the $\rho_\party$ distribution (cf. \FIG\ref{fig:LsMDe}), then under the integral one  has $\eps \ll 1$.
Carrying out the dispersion integral, one obtains a series
\begin{equation}
G_\party(q^2) =  \int_{0}^\infty  \frac{ds \, \rho_\party(s)}{s-q^2-i0}  =
 \frac{1}{q^2-\mE{\party}^2} \Big( \rE{\party} +
\sum_{n \geq 1} {\cal E}_n \Big) \;,
\end{equation}
where
\begin{equation}
{\cal E}_n = \frac{\mom_n}{(q^2 - \mE{\party}^2)^{n}} \;,
\end{equation}
with
\begin{equation}
\label{eq:mom}
\mom_n = \int ds \, \rho_{\party}(s)\, (\mE{\party}^2 - s)^n \;,
\end{equation}
 converging moments  provided  $\rho_{\party}$ has finite support.
For sufficiently Euclidean $q^2$, the hierarchy $|{\cal E}_{n+1}/{\cal E}_n| \ll 1$ is expected to hold, and in practice this may extend to all $q^2 < 0$.
Note that  $\mom_0 = -  \rE{\party}$ which explains the change in sign in the main $q^2$-denominator.   This expansion is  analogous to the multipole expansion in classical electrodynamics, though here it takes place in momentum rather than coordinate space. The analogue of the charge-distribution centre is played by $\mE{\party}$, the approximate
centre of  the $\rho_\party(s)$-distribution.
The  table and figure  in  \APP \ref{app:broad}  show that the
assertions made are true in the \LsM for specific parameter ranges.
See also the discussion in \SEC\ref{sec:G}.

\section{Comparison plots}
\label{app:comparefits}

Figure \ref{fig:mit_ruiz_compare} shows the $D$-form factors of the pion \eqref{eq:pionfunc}  and the nucleon \eqref{eq:Nansatz}  (of the  Euclidean pole fits in \FIG\ref{fig:170mevfits} with $r_{\text{eff}}=0$), compared to  
 to the $n$-pole fits of the original studies~\cite{Hackett:2023rif,Hackett:2023nkr} and to the narrow-resonance 
 approximation in \REFs\cite{Broniowski:2024oyk,Broniowski:2025ctl}.

\begin{figure}[t]
  \centering
  \includegraphics[width=0.49\textwidth]{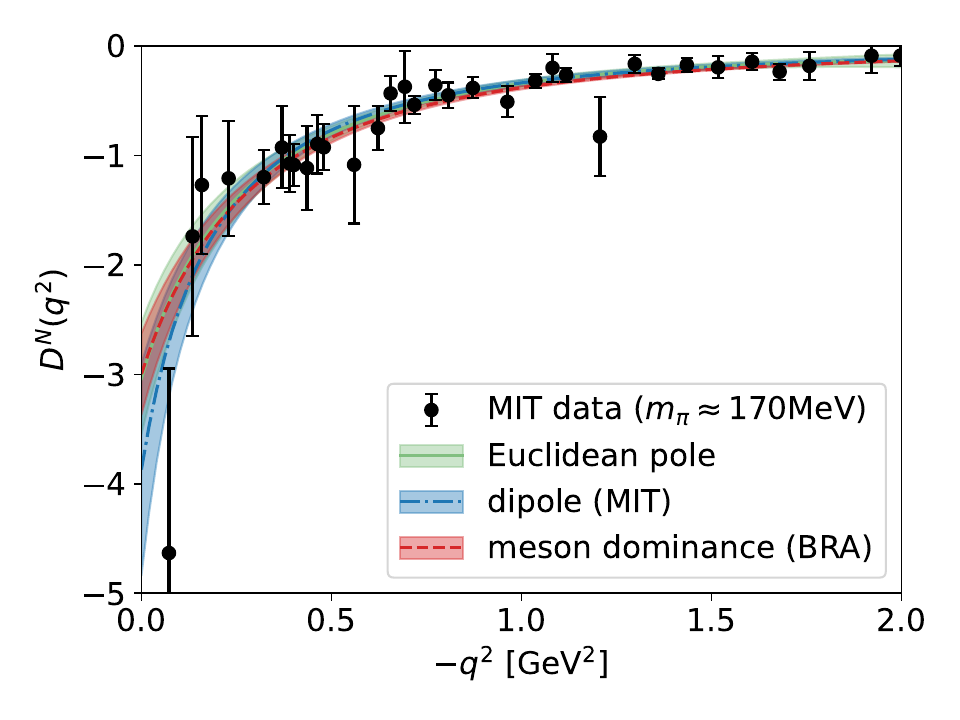}
  \includegraphics[width=0.49\textwidth]{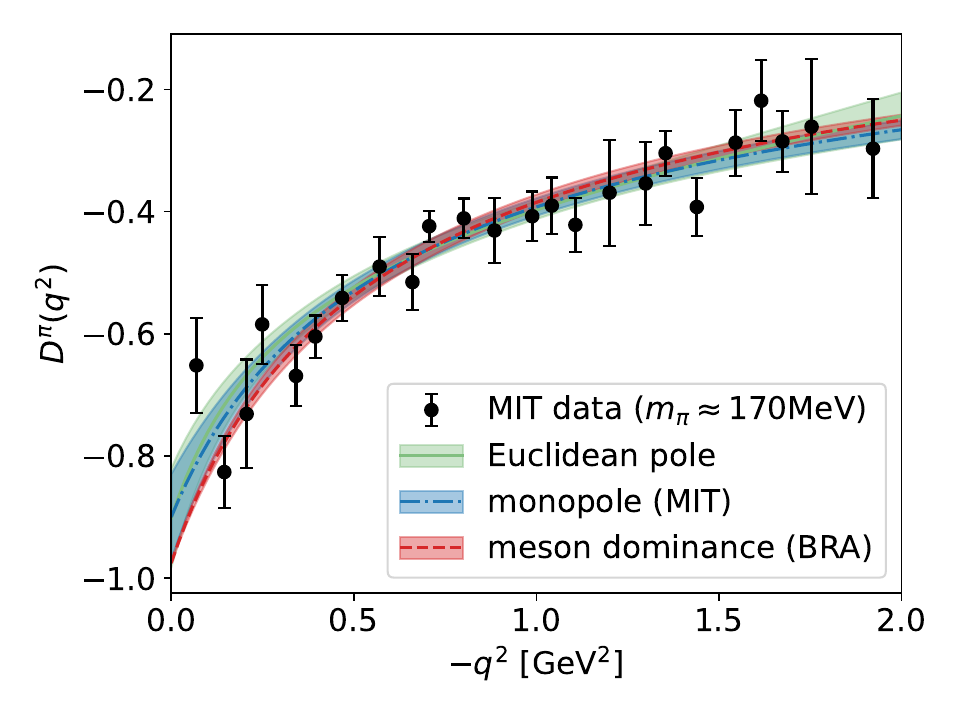}
  \caption{\small The $D$-form factor of the nucleon (left) and pion (right) as a function of the momentum transfer $q^2$. Comparison between our main parametrisation (green) as also shown in \FIG\ref{fig:170mevfits},  i.e., \EQs\eqref{eq:Nansatz} and \eqref{eq:pionfunc} with $\reff=0$, the $n$-pole fits performed in the original paper by Hackett et al.~\cite{Hackett:2023rif,Hackett:2023nkr} (blue) and in the meson dominance approach by Broniowski and Ruiz Arriola~\REFs\cite{Broniowski:2024oyk,Broniowski:2025ctl} (red). The fits are compared to the data of \REFs\cite{Hackett:2023rif,Hackett:2023nkr}.}
  \label{fig:mit_ruiz_compare}
\end{figure}

In both cases agreement at the $1\sig$-level is found, between the three fits, in the entire data-range.  %
However, we wish to emphasise that the aim of this work is not to obtain the best description of this data, but the  physical interpretation in terms of dilaton effective field theory (as discussed in \SEC\ref{sec:GFF}). 

Next we consider  the trace of the energy-momentum tensor $\Theta^{N,\pi}(q^2)$ which is obtained by taking the trace 
of  \EQ\ref{eq:gffdefinition} which are normalised as  $\Theta^{N,\pi}(0) = 2m_{N,\pi}^2$ (for $m_\sig \neq 0$). 
\begin{figure}[t]
  \centering
  \includegraphics[width=0.49\textwidth]{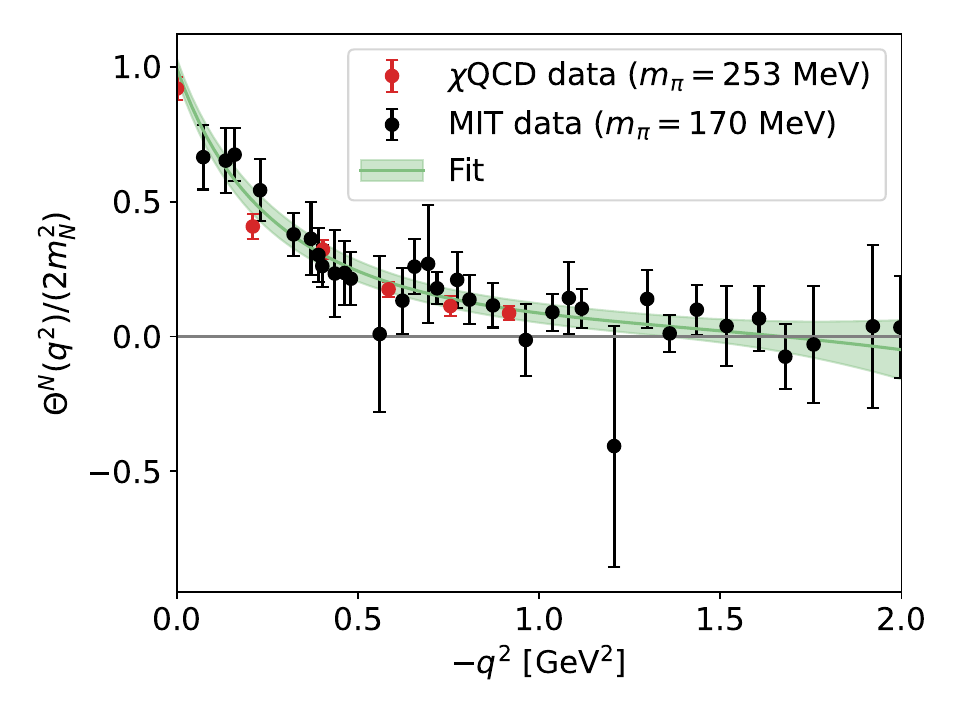}
  \includegraphics[width=0.49\textwidth]{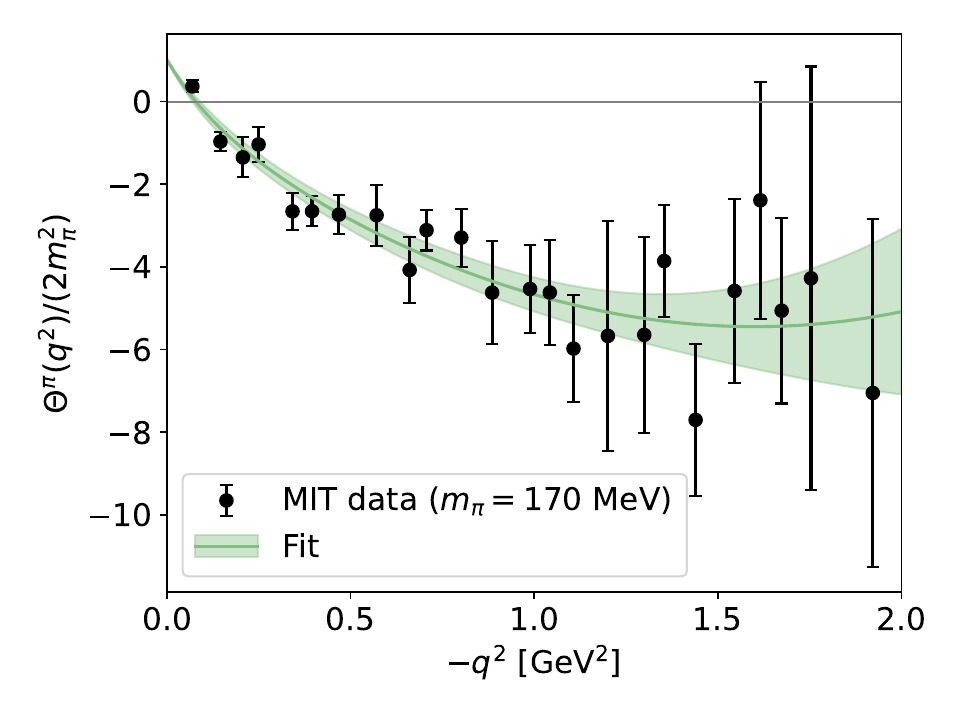}
  \caption{\small The scalar form factor $\GaC(q^2)$ is obtained from the lattice data of \REFs\cite{Hackett:2023nkr,Hackett:2023rif} through $A$, $J$ and $D$ and  uncertainties  by adding in quadrature.
  The green solid line is the fit result with the light band denoting the 68\% c.i., see the main text for more details.
  We show a comparison to the gluonic part of the trace anomaly form factor data by the $\chi$QCD collaboration~\cite{Wang:2024lrm} in red. The plots are normalised such that the function is $1$ for $q^2 =0$. 
  }
  \label{fig:theta}
\end{figure}
In \FIG\ref{fig:theta} we show the fitted $\Theta$-form factors for the nucleon and the pion, from 
the linear combination of the independently fitted  $A$-, $J$- and $D$-form factors.\footnote{Following \REFs\cite{Hackett:2023nkr,Hackett:2023rif}, the $A$- and $J$-form factors are fitted using $n$-pole parametrisations, $F(q^2)=\alpha / (1 - q^2 / \Lambda^2)^n$. Specifically, a monopole parametrisation for the pion and a dipole parametrisation for the nucleon. 
The nucleon dipole fit-parameters  are $\alpha_A=0.99(4)$, $\Lambda_A=1.37(3)$, $\alpha_J=0.50(3)$, $\Lambda_J=-1.51(7)$ with correlations:  $\rho_{\alpha_A,\Lambda_A}=-0.46$, $\rho_{\alpha_D,\Lambda_D}=0.83$.
The pion monopole fit-parameters   are $\alpha_A=1.00(2)$, $\Lambda_A=1.18(2)$ wtith correlations: $\rho_{\alpha_A,\Lambda_A}=-0.14$.} 
Hence,  their uncertainties are added in quadrature and equally so for the   MIT lattice data.
To evaluate the $\Theta$-form factor we take $m_N=903 \MeV$ \cite{private} for the MIT data at $m_\pi=170\MeV$ 
and $m_N=1096\MeV$ (provided with the data) \cite{Wang:2024lrm}  for the $m_\pi=235\MeV$ $\chi \mathrm{QCD}$-data  \cite{Wang:2024lrm}.
 Note that for the  $\chi \mathrm{QCD}$-simulation 
only the field strength part of  the trace anomaly has been considered $\frac{\be}{2g}{G^2}$, omitting 
$m_q (1 + \ga_m) \bar qq$. 
Since the nucleon mass is approximately reproduced for different (pion masses) \cite{Wang:2024lrm}, 
it is clear that this is the lion-share and thus comparison is pragmatically possible.  
We do not compare to the $\chi \mathrm{QCD}$ results for the pion since there the omitted quark contributions 
are too sizeable. 

\bibliographystyle{utphys}
\bibliography{../Dil-refs.bib}

\end{document}